\documentclass[sigconf]{acmart}

\usepackage{booktabs} 
\usepackage{colortbl}
\usepackage{color}
\usepackage{xspace}
\usepackage{url}
\usepackage{balance} 
\usepackage{tabulary}
\usepackage{array}
\usepackage{graphicx}

\usepackage[ruled,lined,linesnumbered,vlined,algo2e]{algorithm2e}
\usepackage{subfig}
\usepackage{listings}
\usepackage{lipsum}
\usepackage{hyperref}
\usepackage{graphicx}
\usepackage{amsmath,amssymb}

\usepackage[T1]{fontenc}
\usepackage{amsmath, amsthm, amssymb}
\usepackage[ansinew]{inputenc}
\usepackage[many]{tcolorbox}

\usepackage{color}
\usepackage{xcolor}
\usepackage{extarrows}
\usepackage{booktabs}
\usepackage{multirow}
\usepackage{colortbl}
\usepackage{tabularx}
\usepackage{enumitem}
\setitemize[0]{leftmargin=10pt}

\let\OLDthebibliography\thebibliography
\renewcommand\thebibliography[1]{
  \OLDthebibliography{#1}
  \setlength{\parskip}{1.6pt}
  \setlength{\itemsep}{0pt plus 0.3ex}
}

\newcommand{\etal}{\hbox{\emph{et al.}}\xspace}
\newcommand{\eg}{\hbox{\emph{e.g.}}\xspace}
\newcommand{\ie}{\hbox{\emph{i.e.}}\xspace}

\newcommand{\etc}{\hbox{\emph{etc}}\xspace}
\definecolor{light-gray}{gray}{0.8}

\newcommand{\quotes}[1]{``#1''}

\makeatletter
\newcommand*\bigcdot{\mathpalette\bigcdot@{1}}
\newcommand*\bigcdot@[2]{\mathbin{\vcenter{\hbox{\scalebox{#2}{$\m@th#1\bullet$}}}}}
\makeatother

\begin{document}
\title{Large-Scale Analysis of Framework-Specific Exceptions in Android Apps}

\author{Lingling Fan\textsuperscript{1,2}, Ting Su\textsuperscript{2$*$\authornote{Ting Su, Lihua Xu and Geguang Pu are the corresponding authors. Lingling Fan and Ting Su contributed equally to this work.}}, Sen Chen\textsuperscript{1,2}, Guozhu Meng\textsuperscript{3,2}}
\author{Yang Liu\textsuperscript{2}, Lihua Xu\textsuperscript{1$*$}, Geguang Pu\textsuperscript{1,4$*$}, Zhendong Su\textsuperscript{5}}
\renewcommand{\authors}{Lingling Fan, Ting Su, Sen Chen, Guozhu Meng, Yang Liu, Lihua Xu, Geguang Pu, Zhendong Su}

\affiliation{\textsuperscript{1}School of Computer Science and Software Engineering, East China Normal University, China} 
\affiliation{\textsuperscript{2}School of Computer Science and Engineering, Nanyang Technological University, Singapore}
\affiliation{\textsuperscript{3}SKLOIS, Institute of Information Engineering, Chinese Academy of Sciences, China}
\affiliation{\textsuperscript{4}Shanghai Key Laboratory of Trustworthy Computing, East China Normal University, China} 
\affiliation{\textsuperscript{5}Department of Computer Science, University of California, Davis, USA}
\email{{ecnujanefan, tsuletgo, ecnuchensen}@gmail.com, {gzmeng,yangliu}@ntu.edu.sg}
\email{lhxu@cs.ecnu.edu.cn, ggpu@sei.ecnu.edu.cn, su@cs.ucdavis.edu}

\renewcommand{\shortauthors}{L.~Fan, T.~Su, S.~Chen, G.~Meng, Y.~Liu, L.~Xu, G.~Pu, and Z.~Su}

\begin{abstract}
Mobile apps have become ubiquitous. For app developers, it is a key priority to ensure their apps' correctness and reliability. However, many apps still suffer from occasional to frequent crashes, weakening their competitive edge.  
Large-scale, deep analyses of the characteristics of real-world app crashes can provide useful insights to guide developers, or help improve testing and analysis tools. However, such studies do not exist --- this paper fills this gap.
Over a four-month long effort, we have collected 16,245 unique exception traces from 2,486 open-source Android apps, and observed that \emph{framework-specific exceptions} account for the majority of these crashes.
We then extensively investigated the 8,243 framework-specific exceptions (which took six person-months): (1) identifying their characteristics (\eg, manifestation locations, common fault categories), (2) evaluating their manifestation via state-of-the-art bug detection techniques, and (3) reviewing their fixes. 
Besides the insights they provide, these findings motivate and enable follow-up research on mobile apps, such as bug detection, fault localization and patch generation.
In addition, to demonstrate the utility of our findings, we have optimized Stoat, a dynamic testing tool, and implemented ExLocator, an exception localization tool, for Android apps.
Stoat is able to quickly uncover three previously-unknown, confirmed/fixed crashes in Gmail and Google+; ExLocator is capable of precisely locating the root causes of identified exceptions in real-world apps. Our substantial dataset is made publicly available to share with and benefit the community. 
	
\end{abstract}

\copyrightyear{2018}
\acmYear{2018}
\setcopyright{acmcopyright}
\acmConference[ICSE '18]{ICSE '18: 40th International Conference on
	Software Engineering }{May 27-June 3, 2018}{Gothenburg, Sweden}
\acmBooktitle{ICSE '18: ICSE '18: 40th International Conference on
	Software Engineering , May 27-June 3, 2018, Gothenburg, Sweden}
\acmPrice{15.00}
\acmDOI{10.1145/3180155.3180222}
\acmISBN{978-1-4503-5638-1/18/05}

\ccsdesc[500]{Software and its engineering~Software testing and debugging}

\keywords{Empirical study, mobile app bugs, testing, static analysis}

\maketitle

\section{Introduction}
\label{sec:intro}

Mobile applications have gained great popularity in recent years.
For example, Google Play, the most popular Android market, has over three million apps, and more than 50,000 apps are continuously published on it each month~\cite{appbrain}.
To ensure competitive edge, app developers and companies strive to deliver high-quality apps.
One of their primary concerns is to prevent fail-stop errors, such as app crashes from occuring in release versions.

Despite the availability of off-the-shelf testing platforms (\eg, Roboelectric~\cite{robolectric}, JUnit~\cite{junit}, Appium~\cite{appium}), and static checking tools (\eg, Lint~\cite{lint}, FindBugs~\cite{findbugs}, SonarQube~\cite{sonarqube})~\cite{KochharTNZL15,LinaresVasquez15}, many released apps still suffer from crashes --- two recent efforts~\cite{MaoHJ16,stoat17} discovered hundreds of previously unknown crashes in popular and well-tested commercial apps. 
Moreover, researchers have contributed a line of work~\cite{AnandNHY12,vanderMerwe2012,MachiryTN13,AzimN13,YangPX13,ChoiNS13,MahmoodMM14,Hao2014,AmalfitanoFTTM15,MaoHJ16,Su16,stoat17,GuC0SDML17,Song2017} to detect app crashes, but none of them have investigated the root causes.
It leaves developers unaware of how to avoid and fix these bugs, and hinders the improvement of bug detection, fault localization~\cite{Sinha09,Jiang10,Mirzaei15,Wu14}, and fixing~\cite{GaoZWXZM15} techniques.
As observed by our investigation on 272,629 issues from 2,174 Android apps hosted on Github and Google Code, developers are unable to resolve nearly 40\% reported crashes,\footnote{Filtered by the keywords \quotes{crash} or \quotes{exception} in their issue descriptions.} which greatly compromises app quality.

This situation underlines the importance of characterizing a large number of diverse real-world app crashes and investigating how to effectively detect and fix them.
However, such a study is difficult and yet to be carried out, which has motivated this work. 

When an app crashes, the Android runtime system will dump an exception trace that provides certain clues of the issue (\eg, the exception type, message, and the invoked methods).
Each exception can be classified into one of three categories --- \emph{application exception}, \emph{framework exception}, and \emph{library exception} --- based on which architecture layer threw the exception.
In particular, our study focuses on framework exceptions, which account for majority of app crashes (affecting over 75\% of the projects), as revealed by our investigation in Section~\ref{sec:rq1}.

We face two key challenges in carrying out the study. The first is the \emph{lack of comprehensive dataset}. To enable crash analysis, we need a comprehensive set of crashes from a large number of real-world apps. Ideally, for each crash, it includes exception trace, buggy source code, bug-triggering inputs, and the patches (if exists). However, to the best of our knowledge, no such dataset is publicly available. Despite open-source project hosting platforms maintain issue repositories, such as Github, our investigation reveals only a small set of crash issues (16\%) are accompanied with exception traces. Among them, even if the issue is closed, it is not necessarily associated with the buggy code version.
The second concerns \emph{difficulties in crash analysis}. Analyzing crashes needs understanding of the application logic as well as the Android framework (or libraries). It is also necessary to cross-validate the root causes (\eg, reproducing crashes, investigating knowledge from developers). However, no reliable tool exists that can facilitate our analysis. 

To overcome these challenges and conduct this study, we made substantial efforts.
We have collected 16,245 unique exception traces from 2,486 open-source Android apps by (1) mining their issue repositories hosted on Github and Google Code; and (2) applying state-of-the-art app testing tools (Monkey~\cite{Monkey}, Sapienz~\cite{MaoHJ16}, and Stoat~\cite{stoat17}) on their recent versions (corresponding to 4,560 executables) to complement the mined data. The whole data collection process took four months. We identified 8,243 unique framework exceptions, and spent nearly six person-months carefully investigating these crashes by examining the source code of apps and the Android framework, fixes from developers, bug reports from testing tools, and technical posts on Stack Overflow. We aim to answer the following research questions:
\begin{itemize}[leftmargin=*]
\item \textbf{\emph{{RQ1}}}: \emph{Compared with other exception categories, are framework exceptions recurring that affect most Android apps? }
\item \textbf{\emph{{RQ2}}}: \emph{What are the common faults made by developers that cause framework exceptions? }
\item \textbf{\emph{{RQ3}}}: \emph{What is the current status of bug detection techniques on detecting framework exceptions? Are they effective?}
\item \textbf{\emph{{RQ4}}}: \emph{How do developers fix framework exceptions? Are there any common practices? What are the difficulties for fixing?} 
\end{itemize}

Through answering the above questions, we aim to characterize Android app crashes (caused by framework exceptions in particular) and provide useful findings to developers as well as researchers. 
For example, our investigation reveals framework exceptions are indeed recurring.
Moreover, they require more fixing efforts (on average 4 days per exception) but have lower issue closing rate (only 53\%) than application exceptions (67\%). Through careful inspection, we distilled 11 common faults that developers are most likely to make, yet have not been well-investigated by previous work~\cite{Hu11,Zaeem14,Coelho15}.
  
We further evaluate the detection abilities of current dynamic testing and static analysis techniques on framework exceptions.
We are surprised to find static analysis tools are almost completely ineffective (only gives correct warnings on 4 out of total 77 exception instances), although there are some plausible ways to improve them. Dynamic testing tools, as expected, can reveal framework exceptions, but still far from effective on certain fault categories. Their testing strategies have a big impact on the detection ability. In addition, we find most exceptions can be fixed by four common practices with small patches (less than 20 code lines), but developers still face several challenges during fixing. 

Our findings enables several follow-up research, \eg, bug detection, fault localization, and patch generation for android apps.
To demonstrate the usefulness of our findings, we have optimized Stoat, a dynamic testing tool, and implemented ExLocator, an exception localization tool, for android apps. The results are promising: Stoat quickly revealed 3 previously unknown bugs in Gmail and Google+; ExLocator is able to precisely localize the root causes of identified exceptions in real apps. 

To summarize, this paper makes the following contributions:
\begin{itemize}
\item To our knowledge, we conducted the first large-scale study to characterize framework-specific exceptions in Android apps, and identified 11 common fault categories that developers are most likely to make. The results provide useful insights for developers and researchers.
\item Our study evaluated the state-of-the-art exception detection techniques, and identified common fixing practices of framework exceptions. The findings shed light on proposing more effective bug detection and fixing techniques.
\item Our findings enable several follow-up research with a large-scale and reusable dataset~\cite{dataset} that contains 16,245 unique exception traces from 2,486 open-source apps. Our prototype tools also demonstrate the usefulness of our findings.
\end{itemize}

\section{Preliminary}
\label{sec:back}

\subsection{Existing Fault Study}
Researchers have investigated Android and Symbian OSes' failures~\cite{Maji10} and Windows Phone app crashes~\cite{Ravindranath14}.
As for the bugs of Android apps, a number of studies exist in different aspects: performance~\cite{Liu:ICSE2014}, energy~\cite{Banerjee16}, fragmentation~\cite{Wei16}, memory leak~\cite{ShahriarNM14,santhanakrishnan2016}, GUI failures~\cite{Adamsen15,AmalfitanoRPF18}, resource usage~\cite{LiuWXC16,Liu16}, API stability~\cite{McDonnell13}, security~\cite{Enck11,meng2017} and \etc.  
However, none of them focus on \emph{functional bugs}, which are also critical to user loyalty and app success.
Our work focuses on this scope.

One of the first attempts at classifying functional bugs is from Hu \etal~\cite{Hu11}. They classify 8 bug types from 10 apps. Other efforts~\cite{Zaeem14,Coelho15}, however, have different goals: Coelho \etal~\cite{Coelho15} analyze exceptions to investigate the bug hazards of exception-handling code (\eg, cross-type exception wrapping), Zaeem \etal~\cite{Zaeem14} study bugs to generate testing oracles for a specific set of bug types.
None of them give a comprehensive analysis, and the validity of their conclusions are unclear.
Therefore, to our knowledge, we are the first to investigate Android app crashes, and give an in-depth analysis.

Our study focuses on the framework-specific exceptions (\emph{framework exception} for short throughout the paper) that can crash apps, \ie, those exceptions thrown from methods defined in the Android framework due to an app's violation of constraints enforced by the framework. Note we do not consider the framework exceptions caused by the bugs of the framework itself. We do not analyze application exceptions (leave this as our future work) and library exceptions (since different apps may use different third-party libraries whose analysis requires other information).

\subsection{Exception Model in Android}
Android apps are implemented in Java, and thus inherit Java's exception model.
Java has three kinds of exceptions.
(1) \texttt{RuntimeException}, the exceptions that are thrown during the normal operation of the Java Virtual Machine when the program violates the semantic constraints (\eg, null-pointer references, divided-by-zero errors).
(2) \texttt{Error}, which represents serious problems that a reasonable application should not try to catch (\eg, \texttt{OutOfMemeoryError}).
(3) \texttt{Checked Exception} (all exceptions except (1) and (2)), these exceptions are required to be declared in a method or constructor's \texttt{throws} clause (statically checked by compilers), and indicate the conditions that a reasonable client program might want to catch. For \texttt{RuntimeException} and \texttt{Error}, the programmers themselves have to handle them at runtime. 

\begin{figure}[h]
\begin{center}
\includegraphics[width=0.45\textwidth]{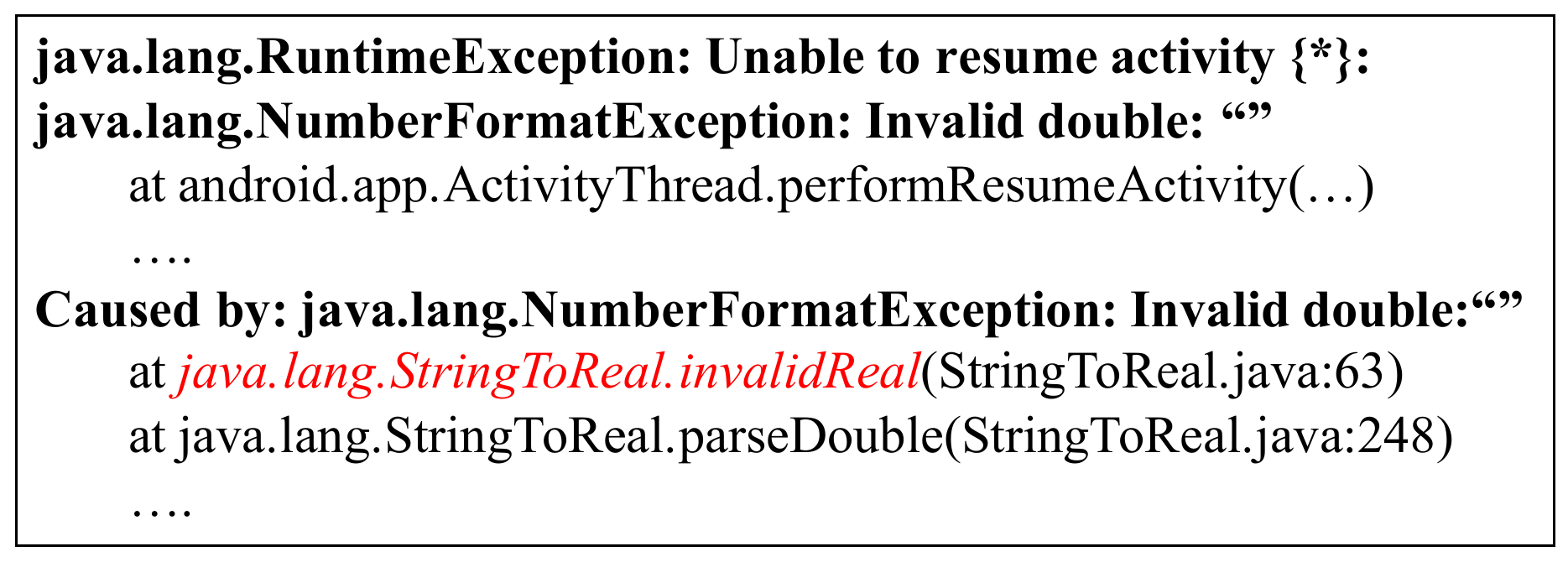}
\end{center}
\vspace{-10pt}
\caption{An example of \texttt{RuntimeException} trace}
\label{fig:exception_trace_example}
\vspace{-10pt}
\end{figure}

Figure~\ref{fig:exception_trace_example} shows an example of \texttt{RuntimeException} trace. The bottom part represents the \emph{root exception}, \ie,  \texttt{NumberFormatException},
which indicates the root cause of this exception. 
Java uses \emph{exception wrapping}, \ie, one exception is caught and wrapped in another (in this case, the \texttt{RuntimeException} of the top part), to propagate exceptions. Note the root exception can be wrapped by multiple exceptions, and the flow from the bottom to the top denotes the order of exception wrappings. 
An \emph{exception signaler} is the method (\texttt{invalidReal} in this case) that throws the exception, which is the first method call under the root exception declaration .

\section{Overview}
\label{sec:overview}

\begin{figure*}[htp]
	\begin{center}
		\includegraphics[width=0.95\textwidth]{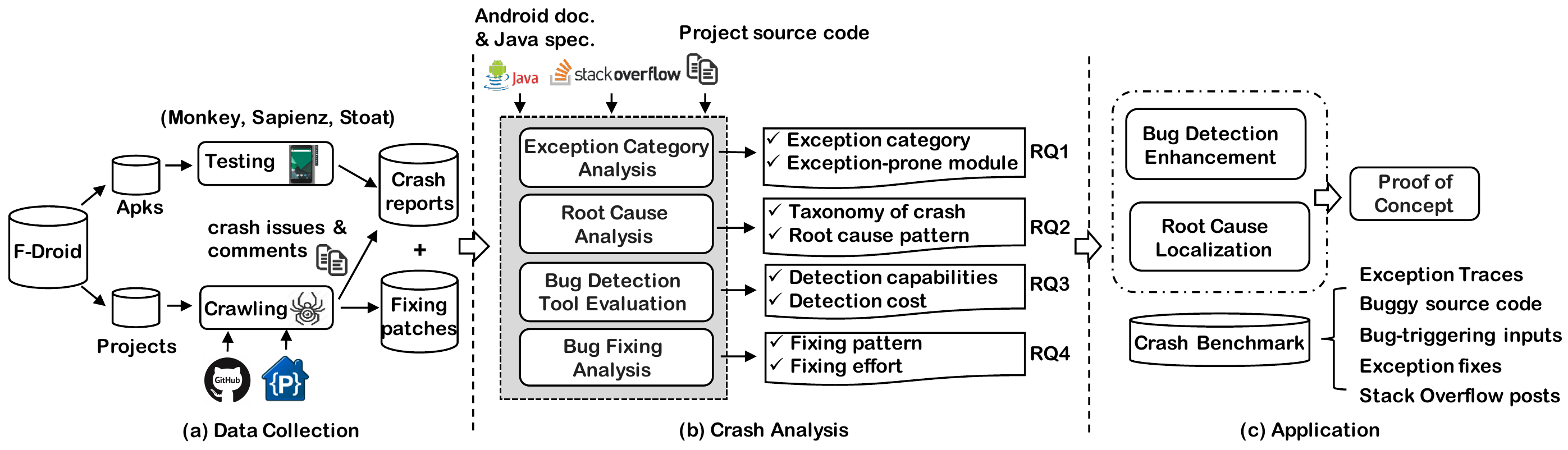}
	\end{center}
	\vspace{-10pt}
	\caption{Overview of our study and its applications}
	\label{fig:study_overview}
\end{figure*}

Figure~\ref{fig:study_overview} shows the overview of our study. We select F-droid~\cite{fdroid} apps as our subjects (Section~\ref{sec:subjects}), and use two methods, \ie, mining issue repositories and applying testing tools, to collect exception traces (Section~\ref{sec:collection}). We investigate exception traces and other resources (\eg, Android documentation, app source code, Stack Overflow posts) to answer RQ1$\sim$RQ4 (Section~\ref{sec:emp}). This study enables several follow-up research detailed in Section~\ref{sec:appl}.

\subsection{App Subjects}
\label{sec:subjects}
We choose F-droid, the largest repository of open-source Android apps, as the source of our study subjects, 
since it has three important characteristics: (1) F-droid contains a large set of apps. At the time of our study, it has more than 2,104 unique apps and 4,560 different app versions, and maintains their metadata (\eg, source code links, release versions).
(2) The apps have diverse categories (\eg, Internet, Personal, Tools), covering different maturity levels of developers, which are the representatives of real-world apps.
(3) All apps are open-source and hosted on Github, Google Code, SourceForge and \etc, which makes it possible for us to access their source code and issue repositories for analysis.

\subsection{Data Collection}
\label{sec:collection}

Table~\ref{table:study_statistics} summarizes the statistics of the collected exception traces. We also collect other data for analysis from Stack Overflow and static analysis tools. The details are explained as follows.

\begin{table}[h]
	\footnotesize
	\centering
	\caption{Statistics of collected crashes}
	\vspace{-10pt}
	\newcommand{\tabincell}[2]{\begin{tabular}{@{}#1@{}}#2\end{tabular}}
	\begin{tabular}{cccc} \hline
		\textbf{Approach} &\textbf{\#Projects} &\textbf{\#Crashes} &\textbf{\#Unique Crashes} \\ \hline \hline
		\tabincell{c}{Hosting Platforms\\\scriptsize{(Github/Google Code)}}  &\tabincell{c}{ 2174\\\scriptsize{(2035/137)} } & \tabincell{c}{7764\\\scriptsize{(7660/104)}} & \tabincell{c}{6588\\\scriptsize{(6494/94)} } \\ 
		\tabincell{c}{Testing Tools\\\scriptsize{(Monkey/Sapienz/Stoat)}}  & \tabincell{c}{ 2104\\\scriptsize{(4560 versions)} }  &\tabincell{c}{13271\\\scriptsize{(3758/4691/4822)}}  & \tabincell{c}{9722\\\scriptsize{(3086/4009/3535)}} \\ \hline
		Total  &2486 \scriptsize{(1792 overlap)} &21035 & 16245\\
		\hline
	\end{tabular}
	\label{table:study_statistics}
	\vspace{-10pt}
\end{table}

\noindent{\emph{\textbf{Github and Google Code}}}. 
We collected exception traces from Github and Google Code since they host over 85\% (2,174/2,549) F-droid apps.
To automate data collection, we implemented a web crawler to automatically crawl the issue repositories of these apps, and collected the issues that contain exception traces. In detail, the crawler visits each issue and its comments to extract valid exception traces. Additionally, it utilizes Github and Google Code APIs to collect project information such as package name, issue id, number of comments, open/closed time. We took about two weeks and successfully scanned 272,629 issues from 2,174 apps, and finally mined 7,764 valid exception traces (6,588 unique) from 583 apps.

\noindent{\emph{\textbf{Automated Testing Tools.}}} 
We set up as follows: (1) We chose three state-of-the-art Android app testing tools with different testing techniques: Google Monkey~\cite{Monkey} (random testing), Sapienz (search-based testing), and Stoat (model-based testing).
(2) We selected all the recent release versions (total 4,560 versions of 2,104 apps, each app has 1$\sim$3 recent release versions) maintained by F-droid as testing subjects. Each tool is configured with default setting and given 3 hours to thoroughly test each version on a single Android emulator. Each emulator is configured with KitKat Android OS (SDK 4.3.1, API level 18). The evaluation is deployed on three physical machines (64-bit Ubuntu/Linux 14.04). Each machine runs 10 emulators in parallel. 
(3) We collect coverage data by Emma~\cite{emma} or JaCoCo~\cite{jacoco} to enable the testing of Sapienz and Stoat.

The evaluation took four months, and finally detected total 13,271 crashes (9,722 unique). In detail, Monkey detected 3,758 crashes (3,086 unique), Sapienz 4,691 crashes (4,009 unique), Stoat 4,822 crashes (3,535 unique).
During testing, we record each exception trace with bug-triggering inputs, screenshots and detection time and \etc, to help our analysis. Further. we find the issue repositories of Github/Google Code only record 545 unique crashes for these recent versions, which accounts for only 5.6\% of those detected by testing tools. This indicates these detected exception traces can effectively complement the mined exceptions.

\noindent{\emph{\textbf{Stack Overflow.}}}
According to exception traces mined from the two sources above, we also collect the most relevant posts on Stack Overflow by searching posts with key word ``Android'', exception types and detailed descriptions. We record information like create time, number of answers, question summary. We mined totally 15,678 posts of various exceptions.

\noindent{\emph{\textbf{Static Analysis Tools.}}}
We also collect data from four state-of-the-art static analysis tools (Lint, PMD, FindBugs, SonarQube), which either serves as a plug-in of Android Studio or supports Android projects. We apply each tool on apps to collect potential bugs, warnings or code smells for in-depth analysis.

\section{Empirical Study}
\label{sec:emp}
\subsection{RQ1: Exception Categories}
\label{sec:rq1}

\noindent{\textbf{Exception Categories}}. 
To investigate app crashes, we group their exception traces into three different categories according to their exception signalers. 
In detail, we refer to Android-18 API documentation~\cite{android_doc} and use the rules below (adopted by prior work~\cite{Coelho15}) to categorize exceptions:
(1) \emph{Application Exception}: the signaler is from the app itself (identified by the app's package name).
(2) \emph{Framework Exception}: the signaler is from the Android framework, \ie, from these packages: ``\texttt{android.*}'', ``\texttt{com.android.*}'', ``\texttt{java.*}'', and ``\texttt{javax.*}''.
(3) \emph{Library Exception}: the signaler is from the core libraries reused by Android (\eg, ``\texttt{org.apache.*}'', ``\texttt{org.json.*}'', ``\texttt{org.w3c.*}'' and \etc) or third-party libraries used by the app.

\begin{table}[h]
	\centering
	\caption{Statistics of the exceptions from Github and Google Code grouped by their signalers (M: Median)}
	\vspace{-10pt}
	\footnotesize
	\newcommand{\tabincell}[2]{\begin{tabular}{@{}#1@{}}#2\end{tabular}}
	\begin{tabular}{lccccccc} \hline
		\textbf{\tabincell{c}{Exception\\ Category}} &\textbf{\#Projects} &\textbf{Occurences} &\textbf{\#Types} &\textbf{\tabincell{c}{Issue \\Duration\\{\tiny{M (Q1/Q3)}}}} & \textbf{\tabincell{c}{Fixing\\Rate}}\\ \hline \hline
		App       & 268 (45.8\%)  & 1552 (23.6\%)  &88 (34\%)   & 2 (0/17) & 67\%  \\  
		Framework & 441 (75.3\%)  & 3350 (50.8\%)  &127 (50\%)  & 4 (1/30) & 53\%  \\  
		Library   & 253 (43.2\%)  & 1686 (25.6\%)  &132 (52\%)   & 3 (1/16) & 57\%  \\  
		\hline
	\end{tabular}
	\label{table:exception_sources}
	\vspace{-10pt}
\end{table}

Table~\ref{table:exception_sources} classifies the exceptions from Github and Google Code according to the above rules, and shows
the number of their affected projects, occurrences, number of exception types, issue durations (the days during the issue was opened and closed), and the fixed issue rate (the percentage of closed issues). From the table, we observe two important facts: 
(1) \emph{Framework exceptions are more pervasive and recurring}. It affects 75.3\% projects, and occupies 50.8\% exceptions.
(2) \emph{Framework exceptions require more fixing effort}. On average, it takes 2 more times effort (see column 5) to fix a framework exception than an application exception

These facts are reasonable. First, most apps heavily use APIs provided by \emph{Android Development Framework} (ADF) to achieve their functionalities. 
ADF enforces various constraints to ensure the correctness of apps, however, if violated, apps may crash and throw exceptions.
Second, fixing application exceptions is relatively easy since developers are familiar with the code logic. However, when fixing framework exceptions, they have to understand and locate the constraints they violate, which usually takes longer.

\noindent{\textbf{Locations of Framework Exception Manifestation}}. 
To further understand framework exceptions, we group them by the class names of their signalers. 
In this way, we get more than 110 groups. To distill our findings, we further group these classes into 17 modules. 
A \emph{module} is used to achieve either one general purpose or stand-alone functionality from the perspective of developers.
We group the classes that manage the Android application model, \eg, \emph{Activities}, \emph{Services}, into \emph{App Management} (corresponding to \texttt{android.app.*}); the classes that manage app data from \emph{content provider} and \emph{SQLite} into \emph{Database} (\texttt{android.database.*}); the classes that provide basic OS services, message passing and inter-process communication into \emph{OS} (\texttt{android.os.*}).
Other modules include \emph{Widget} (UI widgets), \emph{Graphics} (graphics tools that handle UI drawing), \emph{Fragment} (one special kind of activity), \emph{WindowsManager} (manage window display) and \etc.

\begin{figure}[h]
	\begin{center}
		\includegraphics[width=0.4\textwidth]{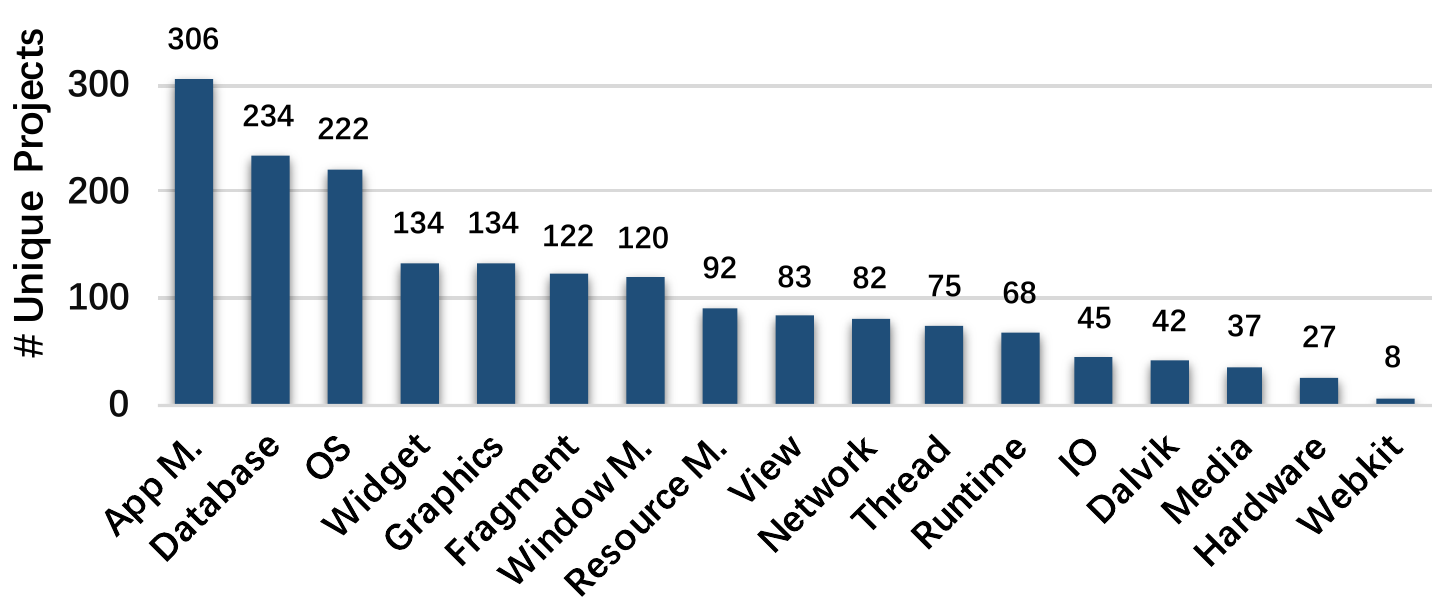}
	\end{center}
	\vspace{-10pt}
	\caption{Exception-proneness of Android modules for framework exceptions (\emph{M.} refers to Management)}
	\label{fig:exception_location}
	\vspace{-10pt}
\end{figure}

Figure~\ref{fig:exception_location} shows the exception-proneness of these modules across all apps.
We find \emph{App Management}, \emph{Database} and \emph{OS} are the top 3 exception-prone modules. In \emph{App Management}, the most common exceptions are \texttt{ActivityNotFound} (caused by no activity is found to handle the given intent) and \texttt{IllegalArgument} (caused by improper registering/unregistering \texttt{Broadcast Receiver} in the activity's callbacks) exceptions. 
Although \texttt{Activity}, \texttt{Broadcast Receiver} and \texttt{Service} are the basic building blocks of apps, surprisingly, developers make the most number of mistakes on them.

As for \emph{Database}, the exceptions of \emph{SQLite} (\eg, \texttt{SQLiteException}, \texttt{SQLiteDatabaseLocked}, \texttt{CursorIndexOutOfBounds}) account for the majority, which reflects the various mistakes of using \emph{SQLite}.
In \emph{OS}, \texttt{SecurityException}, \texttt{IllegalArgument}, \texttt{NullPointer} are the most common ones.
As for the other modules, there are also interesting findings: (1) improper use of \texttt{ListView} with \texttt{Adapter} throws a large number of \texttt{IllegalState} exception (account for 47\%) in \emph{Widget}; (2) improper use of \emph{Bitmap} causes \texttt{OutOfMemoryError} (48\%) in \emph{Graphics};
(3) improper handling callbacks of \emph{Fragment} brings \texttt{IllegalState} (85\%) in \emph{Fragment};
improper showing or dismissing dialogs triggers \texttt{BadTokens} (25\%) in \emph{WindowManager}.

\vspace{2pt}
\noindent\fbox{
	\parbox{0.95\linewidth}{
		\textbf{Answer to RQ1:} \textit{Framework exceptions are pervasive, among which App Management, Database and OS are the three most exception-prone modules for developers.} 
	}
}

\subsection{RQ2: Taxonomy of Framework Exceptions}
\label{sec:rq3}

This section investigates framework exceptions. We classify them into different categories by their root causes. \emph{Root cause}, from the view of developers, is the initial cause of the exception. 

\noindent\emph{\textbf{Exception Buckets}}.
Following the common industrial practice, we group framework exceptions into different buckets. Each bucket contains the exceptions that share the similar root cause.
To achieve this, we use the exception type, message and signaler to approximate the root cause.
For example, the exception in Figure~\ref{fig:exception_trace_example} is labeled as (\texttt{NumberFormatException}, ``\emph{invalid double}'', \texttt{invalidReal}). Finally, we get 2,016 buckets, and find the exceptions from the top 200 buckets have occupied over 80\% of all exceptions. The remaining 20\% buckets have only 5 exceptions or fewer in each of them.
Therefore, we focus on the exceptions of the top 200 buckets.

\noindent\emph{\textbf{Analysis Methods}}. We randomly select a number of exceptions from each bucket, and 
use three complementary resources to facilitate root cause analysis: (1) \emph{Exception-Fix Repository}. We set up a repository that contains pairs of exceptions and their fixes. In particular, (i) from 2,035 Android apps hosted on Github, 
we mined 284 framework exception issues that are closed with corresponding patches. To set up this mapping, we checked each commit message by identifying the keywords \quotes{fix}/\quotes{resolve}/\quotes{close} and the issue id. (ii) We also manually checked the remaining issues to include valid ones that are missed by the keyword rules. We finally got 194 valid issues. We investigate each exception trace and its patch to understand the root causes.
(2) \emph{Exception Instances Repository}. From the 9,722 exceptions detected by testing tools, we filtered out the framework exceptions, and mapped each of them with its exception trace, source code version, bug-triggering inputs and screenshots. When an exception type under analysis is not included or has very few instances in the exception-fix repository, we refer to this repository to facilitate analysis by using available reproducing information. 
(3) \emph{Technical Posts}. For each exception type, we referred to the posts from Stack Overflow collected in Section~\ref{sec:collection} when needing more information from developers and cross-validate our understanding.

\noindent\emph{\textbf{Taxonomy}}.
We analyzed 86 exception types\footnote{After the investigation on a number of \texttt{NullPointerException}s, we find most of them are triggered by null object references. So we did not continue to analyze them.} (covering 84.6\% of all framework exceptions), and finally distilled 11 common faults that developers are most likely to make. Table~\ref{table:root_causes} lists them by the order of closing rate from highest to lowest. We explain them as follows.

\begin{figure}[]
\begin{center}
\includegraphics[width=0.4\textwidth]{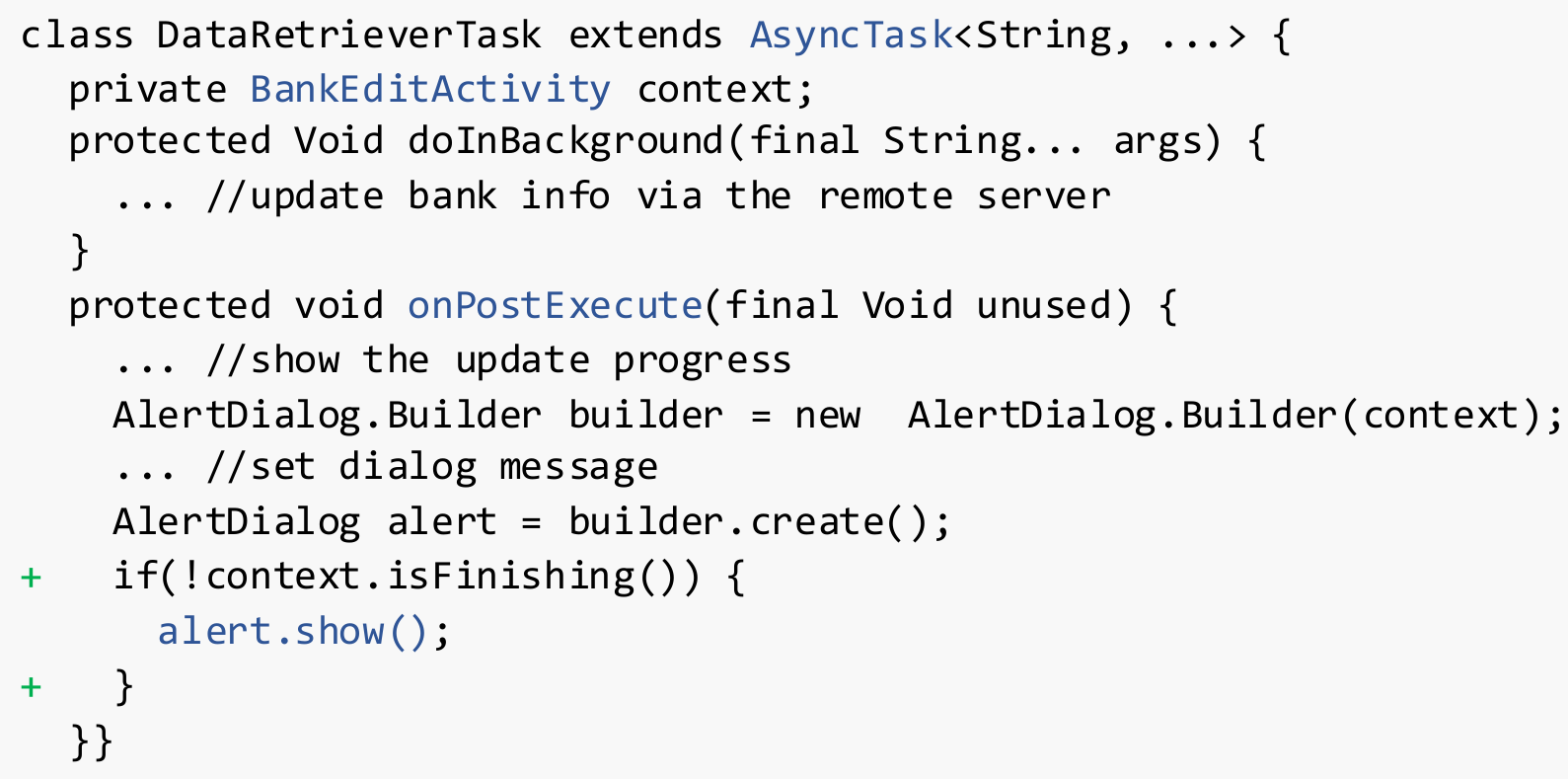}
\end{center}
\vspace{-10pt}
\caption{An Example of Lifecycle Error}
\label{fig:Bankdroid}
\vspace{-10pt}
\end{figure}

\begin{figure}[t]
\begin{center}
\includegraphics[width=0.4\textwidth]{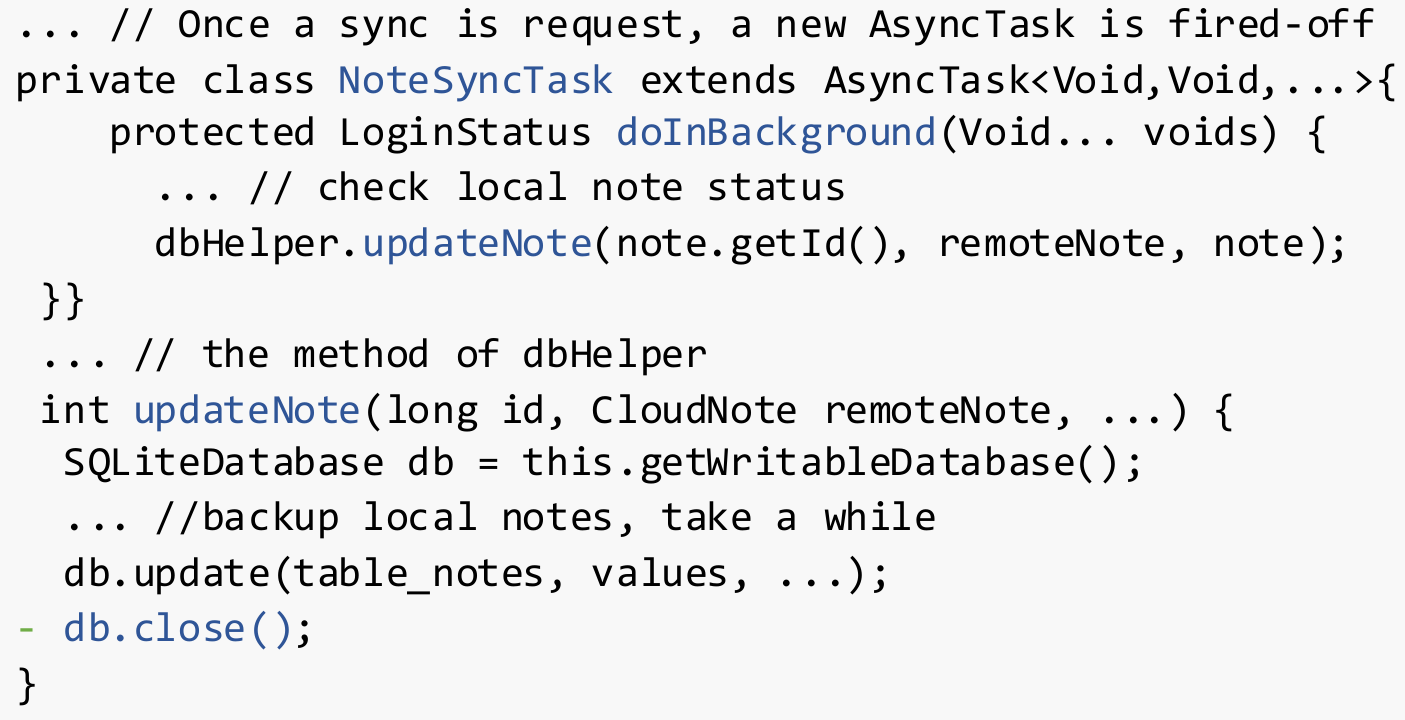}
\end{center}
\vspace{-10pt}
\caption{An Example of Concurrency Error}
\label{fig:Nextcloud}
\vspace{-5mm}
\end{figure}

\noindent \emph{\textbf{$\bigcdot$ Component Lifecycle Error}}. Android apps are comprised of different components. Each component is required to follow a prescribed lifecycle paradigm, which defines how the component is created, used and destroyed~\cite{activity_lifecycle}. For example, \texttt{Activity} provides a core set of six callbacks to allow developers to know its current state.
If developers improperly handle the callbacks or miss state checking before some tasks, the app can be fragile considering the complex environment interplay (\eg, device rotation, network interrupt).
\emph{Bankdroid}~\cite{Bankdroid} (Figure~\ref{fig:Bankdroid}) is an app for providing service of Swedish banks. The app uses a background thread \texttt{DataRetrieverTask} to perform data retrieval, and pops up a dialog when the task is finished. However, if the user edits and updates a bank from \texttt{BankEditActivity} (which starts \texttt{DataRetrieverTask}), during which he presses the back button, the app will crash when the updates finish. The reason is that the developers fail to check \texttt{BankEditActivity}'s state (in this case, \emph{destroyed}) after the task is finished. The bug triggers a \texttt{BadTokenException} and was fixed in revision 8b31cd3~\cite{Bankdroid_revision}. Besides, \texttt{Fragment}~\cite{fragment}, a reusable class implementing a portion of \texttt{Activity}, has much more complex lifecycle. It provides a core set of 12 callbacks to manage its state transition, which makes lifecycle management more challenging, \eg, state loss of \texttt{Fragment}s, attachment loss from its activity.

\begin{figure}[]
\begin{center}
\includegraphics[width=0.4\textwidth]{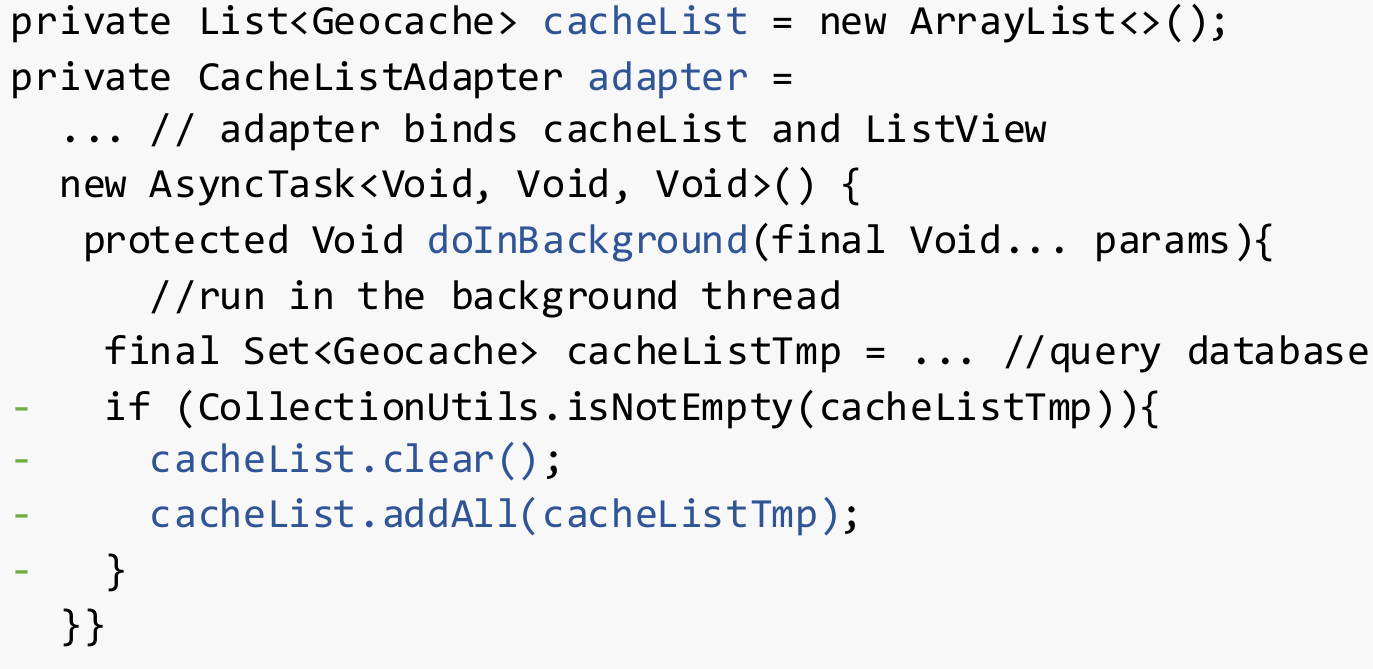}
\end{center}
\vspace{-10pt}
\caption{An Example of UI Update Error}
\label{fig:cgeo}
\vspace{-10pt}
\end{figure}

\noindent \emph{\textbf{$\bigcdot$ Concurrency Error}}. Android system provides such concurrent programming constructs as \texttt{AsyncTask} and \texttt{Thread} to execute intensive tasks.
However, improper handling concurrent tasks may bring data race~\cite{Bielik15} or resource leak~\cite{LiuWXC16}, and even cause app crashes. 
Nextcloud Notes~\cite{Nextcloud} (Figure~\ref{fig:Nextcloud}), a cloud-based notes-taking app that automatically synchronizes local and remote notes, when the app attempts to re-open an already-closed database~\cite{Nextcloud_issue}. The exception can be reproduced by executing these two steps repeatedly: (1) open any note from the list view; (2) close the note as quickly as possible by pressing back-button. The app creates a new \texttt{NoteSyncTask} every time when a note sync is requested, which connects with the remote sever and updates the local database by calling \texttt{updateNote()}. However, when there are multiple update threads, such interleaving may happen and crash the app: \emph{Thread A} is executing the update, and \emph{Thread B} gets the reference of the database; \emph{Thread A} closes the database after the task is finished, and \emph{Thread B} tries to update the closed database. The developers fixed this exception by leaving the database unclosed (since \texttt{SQLiteDatabase} already implemented thread-safe database access mechanism) in revision aa1a972~\cite{Nextcloud_revision}.

\noindent \emph{\textbf{$\bigcdot$ UI Update Error}}. Each Android app owns a UI thread, which is in charge of dispatching events and rendering user interface. To ensure good responsiveness, apps should offload intensive tasks to background threads. However, many developers fail to keep in mind that \emph{Android UI toolkit is not thread-safe and one should not manipulate UI from a background thread}~\cite{thread}. 
\emph{cgeo}~\cite{cego} (Figure~\ref{fig:cgeo}) is a popular full-featured client for geocaching. When refreshing \texttt{cacheList} (\texttt{cacheList} is associated with a \texttt{ListView} via an \texttt{ArrayAdapter}), the developers query the database and substitute this list with new results (via \texttt{clear()} and \texttt{addAll()}) in \texttt{doInbackground}. However, the app crashes when the list is refreshed. The reason is that \texttt{cacheList} is maintained by the UI thread, which internally checks the equality of item counts between \texttt{ListView} and \texttt{cacheList}. But when a background thread touches \texttt{cacheList}, the checking will fail and an exception will be thrown. The developer realized this, and fixed it by moving the refreshing operations into \texttt{onPostExecute}, which instead runs in the UI thread (in revision d6b4e4d~\cite{cego_revision}).

\begin{figure}[]
\begin{center}
\includegraphics[width=0.4\textwidth]{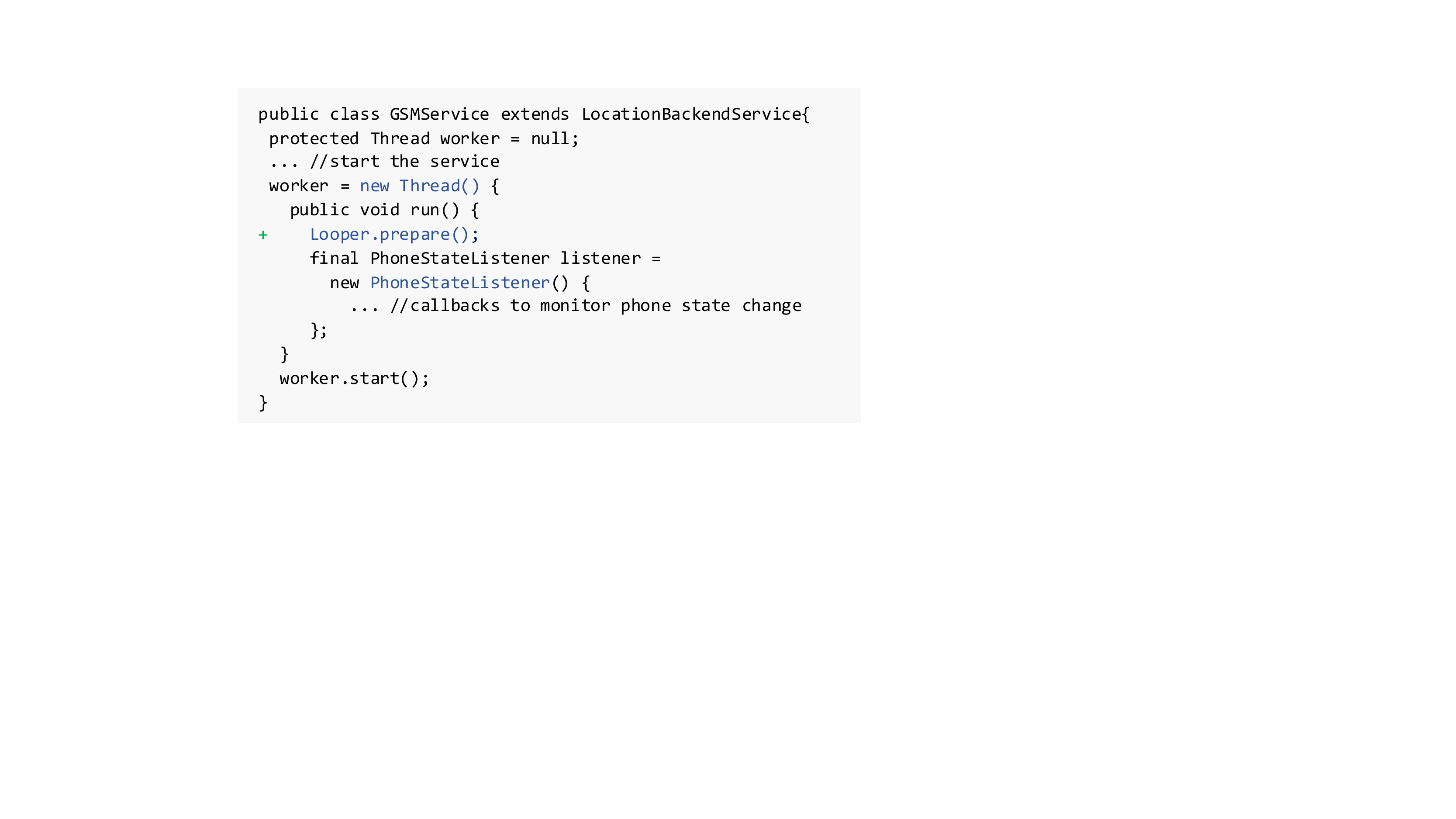}
\end{center}
\vspace{-10pt}
\caption{An Example Violating Framework Constraints}
\label{fig:Local-GSM-Backend}
\vspace{-10pt}
\end{figure}

\begin{table*}[t]
	\centering
	\footnotesize
	\caption{Statistics of 11 common fault categories, and the evaluation results of static analysis tools on them, sorted by \emph{closing rate} in descending order.}
	\label{table:root_causes}
	\begin{tabular}{lcc||c|cccc|l}
		\hline
		\multicolumn{1}{c}{\multirow{2}{*}{Category (Name for short)}} & \multirow{2}{*}{Occurrence} & \multirow{2}{*}{\#S.O. posts}  & \multicolumn{1}{l|}{\multirow{2}{*}{\#Instance}} & \multicolumn{4}{c|}{Static Tools} & \multirow{2}{*}{\begin{tabular}[c]{@{}l@{}} Closing\\ Rate\end{tabular}} \\ \cline{5-8}
		\multicolumn{1}{c}{} &  &  & \multicolumn{1}{l|}{} & Lint & \multicolumn{1}{l}{FindBugs} & \multicolumn{1}{l}{PMD} & \multicolumn{1}{l|}{SonarQube} &  \\ \hline
		API Updates and Compatibility (\textbf{API}) 		& 68& 60  & 7 & - & - & - & - & 93.3\% \\
		XML Layout Error (\textbf{XML}) 					& 122& 246  & 4 & 1 & - & - & - & 93.2\% \\
		API Parameter Error (\textbf{Parameter}) 			& 820& 819  & 6 & - & - & - & - & 88.5\% \\
		Framework Constraint Error (\textbf{Constraint}) 	& 383& 1726  & 12 & 3 & - & - & - & 87.7\% \\
		Others (Java-specific errors)						& 249& 4826  & 10 & - & - & - & - & 86.1\% \\
		Index Error (\textbf{Index}) 						& 950& 218  & 4 & - & - & - & - & 84.1\% \\
		Database Management Error (\textbf{Database}) 		& 128& 61  & 3 & - & - & - & - & 76.8\% \\
		Resource-Not-Found Error (\textbf{Resource}) 		& 1303& 7178  & 5 & - & - & - & - & 75.3\% \\
		UI Update Error (\textbf{UI}) 						& 327& 666  & 3 & - & - & - & - & 75.0\% \\
		Concurrency Error (\textbf{Concurrency}) 			& 372& 263  & 7 & - & - & - & - & 73.5\% \\
		Component Lifecycle Error (\textbf{Lifecycle}) 		& 608& 1065  & 11 & - & - & - & - & 58.8\% \\
		Memory/Hardware Error (\textbf{Memory}) 		 		& 414& 792  & 3 & - & - & - & - & 51.6\% \\ \hline
	\end{tabular}
\end{table*}

\noindent \emph{\textbf{$\bigcdot$ Framework Constraint Error}}. Android framework enforces various constraints for app development. 
For example, \texttt{Handler} is part of Android framework for managing threads, which allows to send and process messages or runnable objects associated with a thread's message queue~\cite{handler}. \emph{Each \texttt{Handler} instance must be associated with a single thread and the message queue of this thread\footnote{A thread by default is not associated with a message queue; to create it, \texttt{Looper\#prepare()} should be called in the thread~\cite{looper}.}}. Otherwise, a runtime exception will be thrown. 
\emph{Local-GSM-Backend}~\cite{LocalGSM} (Figure~\ref{fig:Local-GSM-Backend}), a popular cell-tower based location lookup app, uses a thread \texttt{worker} to monitor the changes of telephony states via \texttt{PhoneStateListener}. However, the developers are unaware that \texttt{PhoneStateListener} internally maintains a \texttt{Handler} instance to deliver messages~\cite{PhoneStateListener}, and thus requires setting up a message loop in \texttt{worker}. They later fixed it by calling \texttt{Looper\#prepare()} (in revision 07e4a759~\cite{LocalGSM_revision}). 
Other constraints include performance consideration (avoid performing network operations in the main UI thread~\cite{Network}, permission consideration (require run-time permission grant for dangerous permissions~\cite{permission} since Android 6.0, otherwise \texttt{SecurityException}) and \etc.

\begin{figure}[]
\begin{center}
\includegraphics[width=0.4\textwidth]{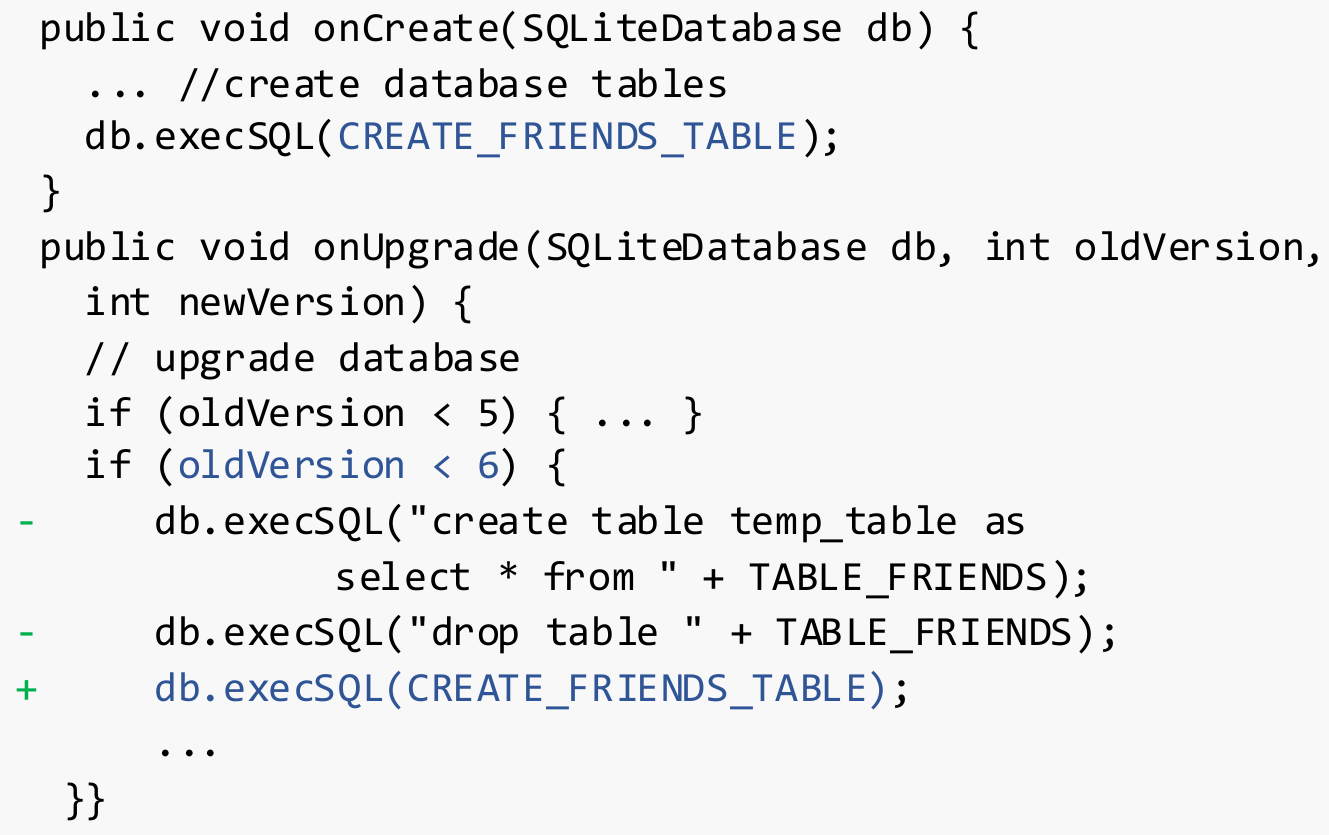}
\end{center}
\vspace{-10pt}
\caption{An Example of Database Management Error}
\label{fig:Atarashii}
\vspace{-10pt}
\end{figure}
\noindent \emph{\textbf{$\bigcdot$ Database Management Error}}. 
Improper manipulating database columns/tables causes many exceptions. Besides this, 
improper data migration for version updates is another major reason.
\emph{Atarashii}~\cite{Atarashii} (Figure~\ref{fig:Atarashii}) is a popular app for managing the reading and watching of anime. When the user upgrades from v1.2 to v1.3, the app crashes once started. The reason is that the callback \texttt{onCreate()} is only called if no old version database file exists, so the new database table \emph{friends} is not successfully created when upgrading. Instead, \texttt{onUpgrade()} is called, it crashes the app because the table \emph{friends} does not exist (fixed in revision b311ec3~\cite{Atarashii_revision}).

\noindent \emph{\textbf{$\bigcdot$ API Updates and Compatibility}}. Android system is evolving fast. API updates and implementation (\eg, SDKs, core libraries) changes can affect the robustness of apps. Device fragmentation~\cite{Wei16} aggravates this problem. 
For example, \texttt{Service} should be started explicitly since Android 5.0; the change of the comparison contract of \texttt{Collections\#sort()}~\cite{collection_issue} since JDK 7 crashes several apps since the developers are unaware of this.

\noindent \emph{\textbf{$\bigcdot$ Memory/Hardware Error}}. Android devices have limited resources (\eg, memory). Improper using of resources may bring exceptions. For example, \texttt{OutOfMemoryError} occurs if loading too large Bitmaps; \texttt{RuntimeException} appears when \texttt{MediaRecorder\#stop()} is called but no valid audio/video data is received.

\noindent \emph{\textbf{$\bigcdot$ XML Design Error}}. Android supports UI design and resource configuration in the form of XML files. Although IDE tools have provided much convenience, mistakes still exist, \eg, misspelling custom UI control names, forgetting to escape special characters (\eg, \quotes{\$}, \quotes{\%}) in string texts, failing to specify correct resources in \texttt{colors.xml} and \texttt{strings.xml}.

\noindent \emph{\textbf{$\bigcdot$ API Parameter Error}}. Developers make this type of mistakes when they fail to consider all possible input contents or formats, and feed malformed inputs as the parameters of APIs. For example, they tend to directly use the results from \texttt{SharedPreference} or database queries without any checking.

\noindent \emph{\textbf{$\bigcdot$ Resource Not Found Error}}. Android apps heavily use external resources (\eg, databases, files, sockets, third-party apps and libraries) to accomplish tasks. Developers make this mistake when they ignore checking their availability before use.

\noindent \emph{\textbf{$\bigcdot$ Indexing Error}}. Indexing error happens when developers access data, \eg, \emph{database}, \emph{string}, and \emph{array}, with a wrong index value. One typical example is the \texttt{CursorIndexOutOfBounds} exception caused by accessing database with incorrect cursor index.

In Table~\ref{table:root_causes}, column 2 and 3, respectively, counts the occurrences of each category and the number of Stack Overflow posts on discussing these faults; column 4 shows the number of distinct exception types of each category (total 75 instances). We find that (1) Besides the ``trivial" errors such as Resource-Not-Found Error, Index Error and API Parameter Error, app developers are more likely to make Android specific errors, \eg, Lifecycle Error, Memory/Hardware Error, Android Framework Constraint Error. (2) developers also discuss more on Android Framework Constraint Error, Lifecycle Error and API Parameter Error.
Additionally, we find existing mutation operators~\cite{DengOAM17,Linares2017} designed for detecting app bugs can cover only a few of these 75 instances.
Deng \etal's 11 operators~\cite{DengOAM17} can only detect 2 instances (the remaining ones detect UI and event handling failures instead of fatal crashes); MDroid+~\cite{Linares2017} proposes 38 operators, but can only cover 8 instances. 

\vspace{2pt}
\noindent\fbox{
	\parbox{0.95\linewidth}{
		\textbf{Answer to RQ2:} \textit{We distilled 11 fault categories that explain why framework exceptions are recurring. Among them, developers make more mistakes on Lifecycle Error, Memory/Hardware Error and Android Framework Constraint Error. Existing mutation operators are inadequate for detecting these errors.} 
	}
}
\vspace{-2pt}

\subsection{RQ3: Auditing Bug Detection Tools}
\label{sec:rq4}

Dynamic testing and static analysis are the two main avenues to help detect software bugs.
This section investigates the detection abilities of these two techniques on framework exceptions (categorized in Section~\ref{sec:rq3}). 
In particular, we select three state-of-the-art testing tools, \ie, Monkey, Sapienz, and Stoat;
and four static analysis tools widely used by android developers~\cite{LinaresVasquez15}, \ie, Lint, FindBugs, PMD, and SonarQube.
Lint, developed by Google, detects code structural problems, and scans for android-specific bugs~\cite{lintchecks}.
PMD uses defect patterns to detect bad coding practices. 
FindBugs, provided as a plugin in Android Studio, also enforces various checking rules, and adopts control- and data-flow analysis to scan potential bugs (\eg, null-pointer dereferences). 
SonarQube is a continuous code quality platform that provides bug reports for suspicious code.

\noindent\emph{\textbf{Static Analysis Tools}}.
We randomly select 75 distinct exception instances (corresponding to column 4 in Table~\ref{table:root_causes}) from Github that cover all manifestations of root faults, and checkout the corresponding buggy code version to investigate how many of them can be detected by static analysis tools. 
Our investigation finds static tools specialize in detecting bad practices, code smells, and potential bugs that may lead to severe errors, but with a mass of false alarms.

As shown in Table~\ref{table:root_causes}, FindBugs, PMD, and SonarQube fail to report any warnings on these bugs. Lint only identifies 4 out of 75 bugs, which include one XML error (the resource file ``string.xml'' contains an illegal character ``\$'') and three framework constraint errors (duplicate resource ids within a layout file; Fragment cannot be instantiated; using the wrong AppCompat method). 
In addition, although these tools claim to support android projects, we have not found any android-specific rules in FindBugs and SonarQube, and only three android rules~\cite{pmdrules} in PMD. Lint defines 281 android rules~\cite{lintchecks} but detects only a few bugs.
Therefore, the current static analysis tools focus more on basic Java defects, and much less effective in detecting framework exceptions of Android apps.

\begin{figure}
  \centering
  \includegraphics[width=0.5\textwidth]{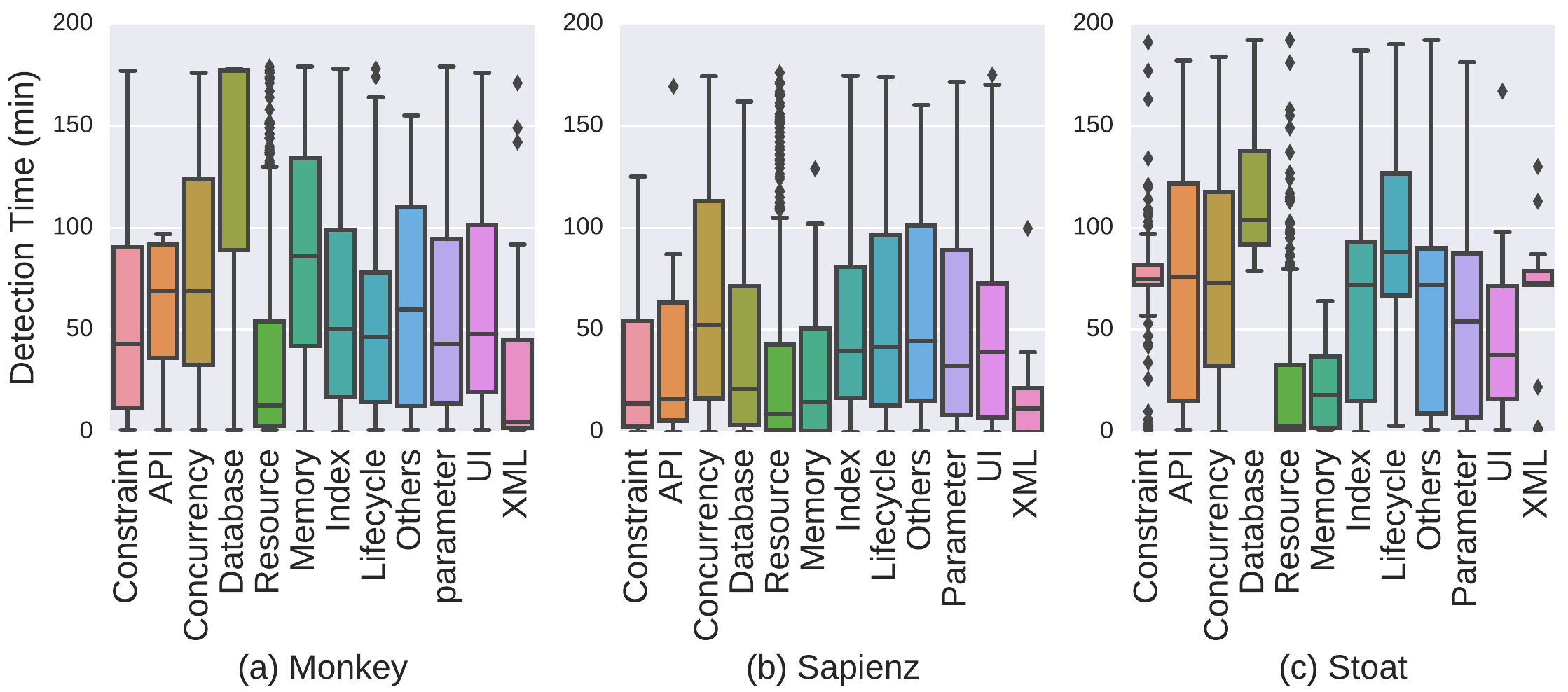}
  \vspace{-10pt}
  \caption{Detection time of exceptions by each tool}
  \label{fig:time}
 \vspace{-10pt}
\end{figure}

\begin{figure}
  \centering
  \includegraphics[width=0.5\textwidth]{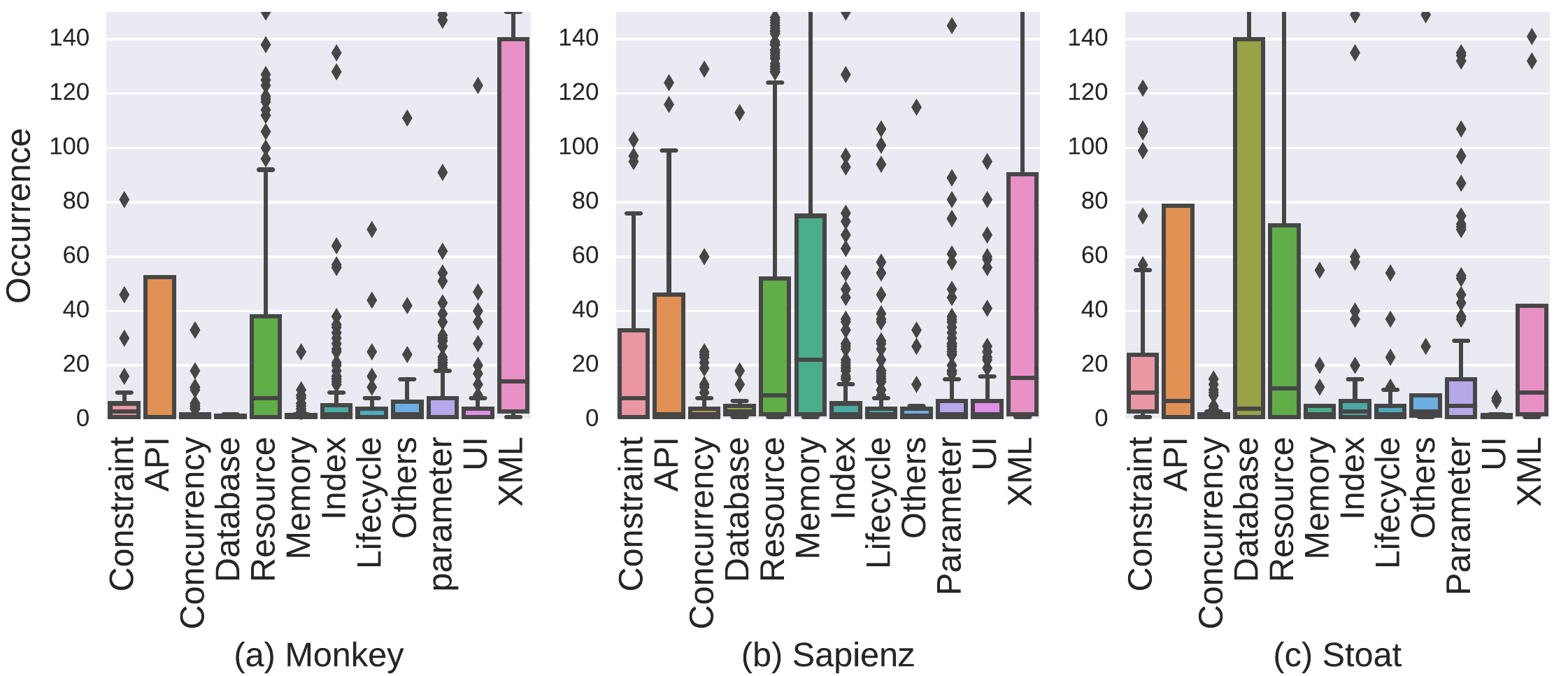}
  \vspace{-10pt}
  \caption{Occurrences of exceptions by each tool}
  \label{dynamic:occurrence}
  \vspace{-10pt}
\end{figure}

\noindent\emph{\textbf{Dynamic Testing Tool}}.
We apply testing tools on each app (total 2,104) with the same configuration in Section~\ref{sec:collection}.
As we observed, they can detect many framework exceptions. To understand their abilities, we use two metrics\footnote{We have not presented the results of trace length, since we find the three tools cannot dump the exact trace that causes a crash. Instead, they output the whole trace, which cannot reflect their detection abilities.}.
(1) \emph{detection time} (the time to detect an exception). Since one exception may be found multiple times, we use the time of its first occurrence. (2) \emph{Occurrences} (how many times an exception is detected during a specified duration).  
Figure~\ref{fig:time} and Figure~\ref{dynamic:occurrence}, respectively, show the detection time and occurrences of exceptions by each tool grouped by the fault categories. 

Figure~\ref{fig:time} shows concurrency errors are hard to detect for all three tools (requiring longer time). But for other fault categories, the time varies on different tools. For example, 
Sapienz is better at database errors (since Sapienz implements a strategy, \ie, fill strings in \texttt{EditText}s, and then click ``OK'' instead of ``Cancel'' to maximize code coverage, which is more likely to trigger database operations); Monkey and Sapienz are better at lifecycle errors (since both of them emit events very quickly without waiting the previous ones to take effect, \eg, open and quickly close an activity without waiting the activity finishes its task).
Figure~\ref{dynamic:occurrence} shows it is easy for three tools to detect API compatibility, Resource-Not-Found and XML errors since the occurrences of these errors are much more than those of the others. 
But for other categories, \eg, Concurrency, Lifecyle, Memory, UI update errors, all of three tools are far from effective regardless of their testing strategies. The main reason is that these errors contain non-determinism (interact with threads).

After an in-depth inspection, we find that some Database errors are hard to trigger because the app has to construct an appropriate database state (\eg, create a table or insert a row, and fill in specific data) as the precondition of the bug, which may take a long time. As for Framework Constraint errors, some exceptions require special environment interplay. For example, \texttt{InstantiationException} of Fragment can only be triggered when a Fragment (without an empty constructor) is destroyed and recreated. To achieve this, a testing tool needs to change device rotation at an appropriate timing (when the target Fragment is on the screen), or pause and stop the app by switching to another one, and stay there for a long time (let Android OS kill the app), and then return back to the app. Concurrency bugs are hard to trigger since they usually need right timings of events.

\vspace{1mm}
\noindent\fbox{
	\parbox{0.95\linewidth}{
		\textbf{Answer to RQ3:} \textit{Existing static analysis tools are ineffective in detecting framework exceptions. Dynamic testing tools are still far from effective in detecting database, framework constraint and concurrency errors.} 
	}
}

\subsection{RQ4: Fixing Patterns and Efforts}
\label{sec:rq5}

This section uses the exception-fix repository constructed in RQ2 (194 instances) to investigate the common practices of developers to fix framework exceptions. We categorize their fixing strategies by (1) the types of code modifications (e.g., modify conditions, reorganize/move code, tweak implementations); (2) the issue comments and patch descriptions. We finally summarized 4 common fix patterns, which can resolve over 80\% of the issues.

\begin{figure}[]
\begin{center}
\includegraphics[width=0.4\textwidth]{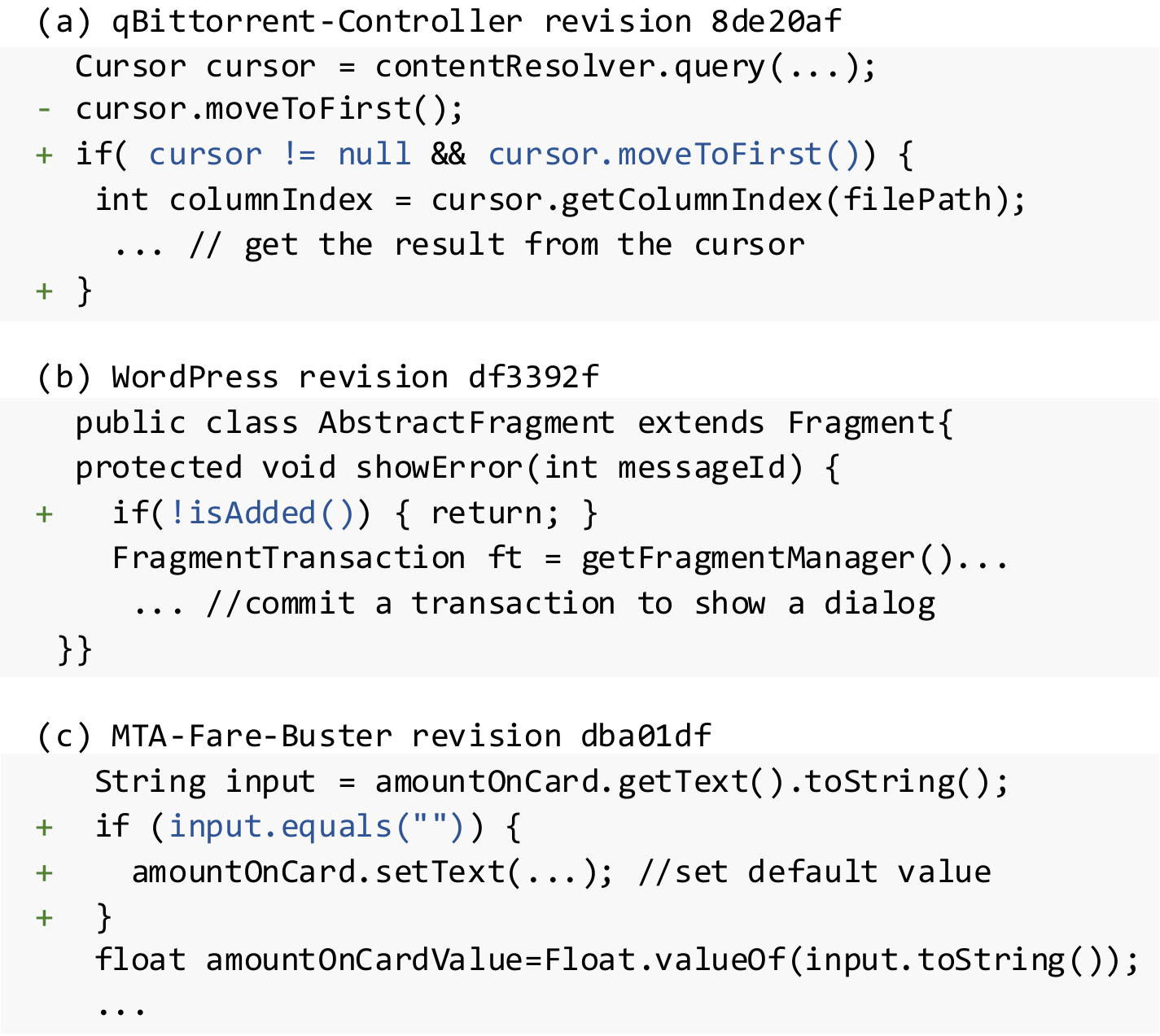}
\end{center}
\vspace{-10pt}
\caption{Example fixes by adding conditional checks}
\label{fig:example_fixes_conditional_checks}
\vspace{-10pt}
\end{figure}

\begin{figure}[]
\begin{center}
\includegraphics[width=0.4\textwidth]{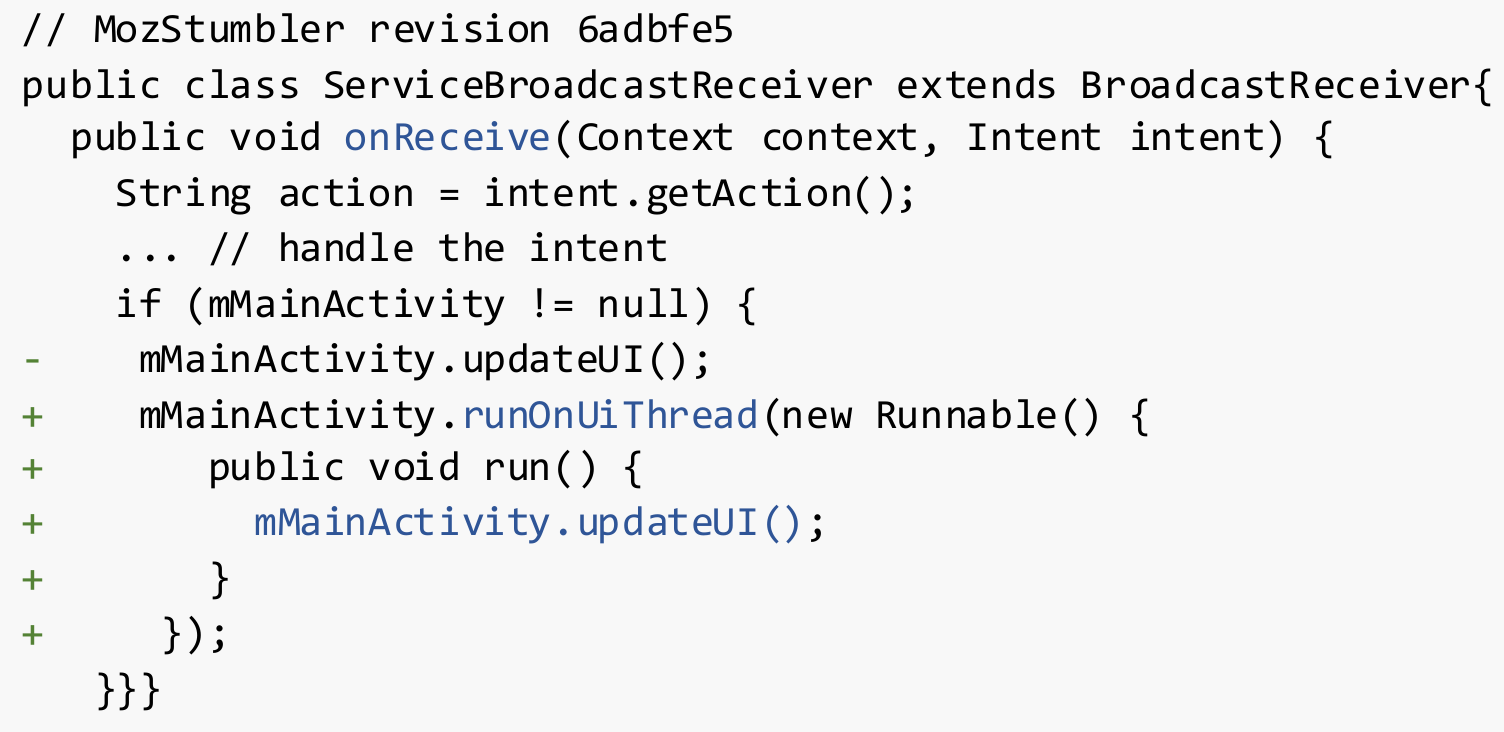}
\end{center}
\vspace{-10pt}
\caption{Example fixes by moving code into correct thread}
\label{fig:example_fixes_correct_thread}
\vspace{-10pt}
\end{figure}

\noindent \emph{\textbf{$\bigcdot$ Refine Conditional Checks}}. Missing checks on API parameters, activity states, index values, database/SDK versions, external resources can introduce unexpected exceptions. Developers usually fix them via adding appropriate conditional checks. For example, Figure~\ref{fig:example_fixes_conditional_checks} (a) checks cursor index to fix \texttt{CursorIndexOutOfBound}, Figure~\ref{fig:example_fixes_conditional_checks} (b) checks the state of the activity attached by a Fragment to fix \texttt{IllegalState}, and Figure~\ref{fig:example_fixes_conditional_checks} (c) checks the input of an \texttt{EditText} to fix \texttt{NumberFormat}.
We find most of exceptions from \emph{Parameter Error}, \emph{Indexing Error}, \emph{Resource Error}, \emph{Lifecycle Error}, and \emph{API Error} can be fixed by this strategy. 

\noindent \emph{\textbf{$\bigcdot$ Move Code into Correct Thread}}. Messing up UI and background threads may incur severe exceptions. The common practice to fix such problems is to move related code into correct threads. Figure~\ref{fig:example_fixes_correct_thread} fixes \texttt{CalledFromWrongThread} by moving the code of modifying UI widgets back to the UI thread (via \texttt{Activity\#runOnUiThread()}) that creates them. Similar fixes include moving the showings of \texttt{Toast} or \texttt{AlertDialog} into the UI thread instead of the background thread since they can only be processed in the \texttt{Looper} of the UI thread~\cite{Toast_exception,AlertDiaglog_exception}. Additionally, moving extensive tasks (\eg,  network access, database query) into background thread can resolve such performance exceptions as \texttt{NetworkOnMainThread} and ``Application Not Responding" (ANR)~\cite{ANR}.

\noindent \emph{\textbf{$\bigcdot$ Work in Right Callbacks}}. Inappropriate handling lifecycle callbacks of app components (\eg, \texttt{Activity}, \texttt{Fragment}, \texttt{Service}) can severely affect the robustness of apps. The common practice to fix such problems is to work in the right callback. For example, in \texttt{Activity}, putting \texttt{BroadcastReceiver}'s register and unregister into \texttt{onStart()} and \texttt{OnStop()} \emph{or} \texttt{onResume()} and \texttt{OnPause()} can avoid \texttt{IllegalArgument}; and committing a \texttt{FragmentTransaction} before the activity's state has been saved (\ie, before the callback \texttt{onSaveInstanceState()}) can avoid state loss exception~\cite{state_loss,Shan16}.

\noindent \emph{\textbf{$\bigcdot$ Adjust Implementation Choices}}. To resolve other exceptions, developers have to adjust the implementation or do code refactoring. For example, to fix \texttt{OutOfMemory} caused by loading Bitmap, the common practice is to optimize memory usage by resizing the original bitmap~\cite{bitmap_memory}; to fix data race exceptions, the common practice is to adopt mutex locks (\eg, add \emph{synchronized} to allow the execution of only one active thread) or back up the shared data~\cite{MPDroid_issue}.

\begin{figure}[t]
\begin{center}
\includegraphics[width=0.48\textwidth]{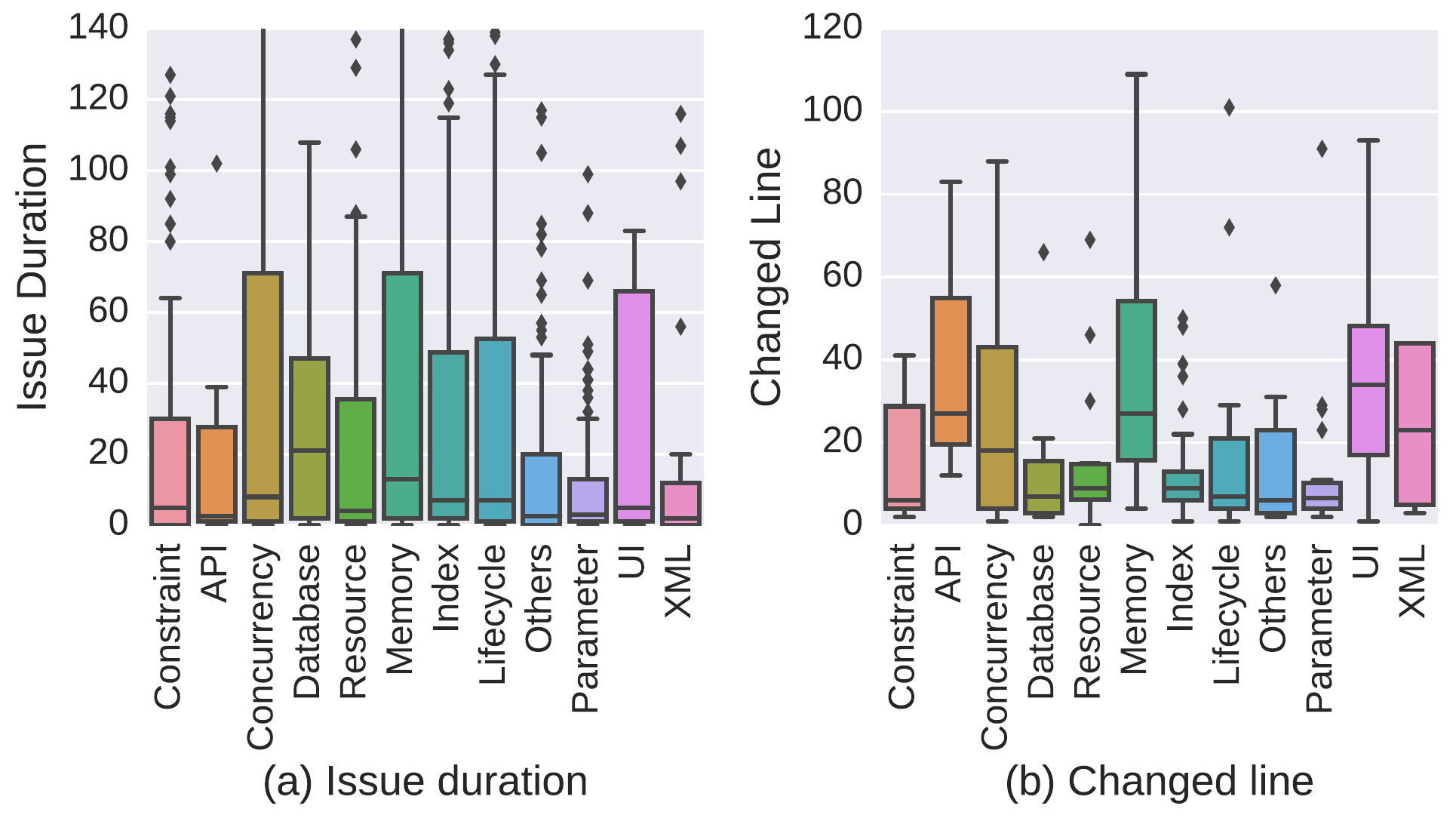}
\end{center}
\vspace{-10pt}
\caption{Fixing Effort}
\label{fig:fix_efforts}
\vspace{-10pt}
\end{figure}

To further understand the characteristics of developer fixes, we group these issues by their root causes, and compute three metrics: 
(1) \emph{Issue Duration}, which indicates how long the developers took to fix the issue (Figure~\ref{fig:fix_efforts}(a)); (2) \emph{Number of Changed Code Lines}, \ie, the number of code lines\footnote{To reduce ``noises", we exclude comment lines (\eg, ``//...''), annotation lines (\eg, ``@Override''), unrelated code changes (\eg, ``import *.*'', the code for new features).} the developers changed to fix this issue (Figure~\ref{fig:fix_efforts}(b)); and (3) \emph{Issue Closing Rate}, \ie, how many issues have been closed (the last column in Table~\ref{table:root_causes}).
We can see that the fixes for \emph{Parameter Error}, \emph{Indexing Error}, \emph{Resource Error}, and \emph{Database Error} require fewer code changes (most patches are less than 20 lines). Because most of them can be fixed by refining conditional checks. We also note \emph{UI Error}, \emph{Concurrency Error}, and \emph{Memory/Hardware Error} require larger code patches. 

Further, by investigating the discussions and comments of developers when fixing, we find three important reasons that reveal the difficulties they face. 

\noindent \emph{\textbf{$\bigcdot$ Difficulty of Reproducing and Validation}}. One main difficulty is how to reproduce exceptions and validate the correctness of fixes~\cite{MoranVBVP16}. Most users do not report complete reproducing steps/inputs and other necessary information (\eg, exception trace, device model, code version) to developers. Even if the exception trace is provided, reproducing such exceptions as non-deterministic ones (\eg, concurrency errors) is rather difficult. In such cases, after fixing the issue, they choose to leave it for the app users to validate before closing the issue. As shown in Figure~\ref{fig:fix_efforts} and Table~\ref{table:root_causes}, concurrency errors have longer issue durations and lower fixing rate.

\noindent \emph{\textbf{$\bigcdot$ Inexperience with Android System}}. A good understanding of Android system is essential to correctly fix exceptions. As the closing rates in Table~\ref{table:root_causes} indicate, developers are more confused by Memory/Hardware Error, Lifecycle Error, Concurrency Error, and UI Error. We find some developers use simple \emph{try-catch} or compromising ways (\eg, use \texttt{commitAllowingStateLoss} to allow activity state loss) as workarounds. However, such fixes are often fragile.

\noindent \emph{\textbf{$\bigcdot$ Fast Evolving APIs and Features}}. Android is evolving fast. As reported, on average, 115 API updates occur each month~\cite{McDonnell13}. Moreover, feature changes are continuously introduced. However, these updates or changes may make apps fragile when the platform they are deployed is different from the one they were built; and the developers are confused when such issues appear. For example, Android 6.0 introduces runtime permission grant --- If an app uses dangerous permissions, developers have to get permissions from users at runtime. However, we find several developers choose to delay the fixing since they have not fully understand this new feature.

\vspace{2pt}
\noindent\fbox{
	\parbox{0.95\linewidth}{
		\textbf{Answer to RQ4:} \textit{Refining conditional checks, using correct thread types, working in the right callbacks, adjusting implementation choices are the 4 common fix practices. 
		Memory/Hardware, Lifecycle, Concurrency, and UI update Error are more difficult to fix.}
		} 
	}

\subsection{Discussion}
Through this study, we find:
(1) Besides the trivial errors, developers are most likely to introduce Lifecycle, Memory/Hardware, Concurrency, and UI update errors, which
requires more fixing efforts.
(2) Bug detection tools need more enhancement. Static analysis tools could integrate new rules especially for UI update, Lifecycle, Framework Constraint errors. Testing tools could integrate specific testing strategies to detect these errors.
(3) To counter framework exceptions, developers should gain more understanding on Android system; different supporting tools should be developed to reproduce exceptions for debugging, locate their root causes for fixing, and check API compatibility across different SDKs.

Linares-V\'{a}squez \etal~\cite{Linares2017} also investigated android app bugs very recently, but our study significantly differs from theirs.
We focuses on framework exceptions and give a comprehensive, deep analysis, including exception manifestations, root causes, 
abilities of existing bug analysis tools, and fixing practices. While they focus on designing mutation operators from existing bugs, and their 38 operators only cover 8 out of 75 instances distilled by our study. We believe our results can further improve existing mutation operators.

The validity of our study may be subject to some threats.
(1) \emph{Representativeness of the subjects}. To counter this, we collected all the subjects (total 2486 apps at the time of our study) from F-Droid, which the largest database of open-source apps, and covers diverse app categories. We believe these subjects are the representatives of real-world apps.
(2) \emph{Comprehensiveness of app exceptions}. To collect a comprehensive set of exception traces, we mine from Github and Google Code; and apply testing tools, which leads to total 16,245 exceptions. 
To our knowledge, this is the largest study for analyzing Android app exceptions.
(3) \emph{Completeness/Correctness of exception analysis}. For completeness, (i) we investigated 8,243 framework exceptions, and carefully inspected all common exception types. (ii) We surveyed previous work~\cite{Hu11,AmalfitanoFTCM12,MachiryTN13,MahmoodMM14,Zaeem14,HuYTY14,Bielik15,AmalfitanoFTTM15,Coelho15,MaoHJ16,lifecycle16,stoat17,model16} that reported exceptions, and observed all exception types and patterns were covered by our study. For correctness, we cross-validated our analysis on each exception type, and also referred to the patches from developers and Stack Overflow posts to validate our analysis. The whole dataset is also made publicly available.

\section{Applications of Our Study}
\label{sec:appl}
This section discusses the follow-up research motivated by our findings, and also demonstrates usefulness by two prototype tools.

\subsection{Benchmarking Android App Exceptions}
Our study outputs a large and comprehensive dataset of Android app exceptions (especially for framework exceptions), which includes total 16,245 unique app exceptions and 8,243 unique framework exceptions. Each exception is accompanied with buggy code version, exception trace, error category, and possible fixes.
We believe this dataset can (1) provide an effective and reliable basis for comparing dynamic/static analysis tools; (2) enable the research on investigating fault localization techniques and give a large set of exceptions as benchmarks; and (3) enable patch generation by comparing the exceptions and their fixes.

\subsection{Improving Exception Detection}
Dynamic testing and static analysis are the two avenues to detect faults.
However, more improvements should be made on both sides. 

\noindent{\emph{\textbf{Dynamic Testing}}}. Enhancing testing tools to detect specific errors is very important. For example,
(1) \emph{Generate meaningful as well as corner-case inputs to reveal parameter errors}. We find random strings with specific formats or characters are very likely to reveal unexpected crashes. For instance, Monkey detects more \texttt{SQLiteException}s than the other tools since it can generate strings with special characters like \quotes{"} and \quotes{\%} by randomly hitting the keyboard. When these strings are used in SQL statements, they can fail SQL queries without escaping.
(2) \emph{Enforce environment interplay to reveal lifecycle, concurrency and UI update errors}. We find some special actions, \eg, change device orientations, start an activity and quickly return back without waiting it to finish, put the app at background for a long time (by calling another app) and return back to it again, can affect an app's internal states and its component lifecycle. Therefore, these actions can be interleaved with normal UI actions to effectively check robustness.
(3) \emph{Consider different app and SDK versions to detect regression errors}. We find app updates may introduce unexpected errors. For example, as shown in Figure~\ref{fig:Atarashii}, the changes of database scheme can crash the new version since the developers have not carefully managed database migration from the old version. (4) \emph{More advanced testing criteria}~\cite{Borges17,Su17} are desired.

\noindent{\emph{\textbf{Static Analysis}}}. 
Incorporating new checking rules into static analysis tools to enhance their abilities is highly valuable.
Through our study, we find it is feasible to check some framework exceptions, especially for framework constraint, lifecycle and UI update errors.
For example, to warn the potential crash in Figure~\ref{fig:Local-GSM-Backend}, static analysis can check whether the task running in the thread uses \texttt{Handler} to dispatch messages, if it uses, \texttt{Looper\#prepare()} must be called at the beginning of \texttt{Thread\#run()}; to warn the potential crash in Figure~\ref{fig:Bankdroid}, static analysis can check whether there is appropriate checking on activity state before showing a dialog from a background thread. In fact, there is already some initial work~\cite{lifecycle16} that implements lifecycle checking on Lint.

\noindent{\emph{\textbf{Demonstration of Usefulness}}}. 
We enhanced Stoat~\cite{stoat17} with two strategies: (1) randomly generate inputs with 5 specific formats (\eg, empty string, lengthy string, null) or characters (\eg, \quotes{"}, \quotes{\%}) to fill in \texttt{EditText}s or \texttt{Intent}'s fields; (2) randomly inject 3 types of special actions mentioned above into normal UI actions. We applied Stoat on dozens of most popular apps (\eg, Facebook, Gmail, Google+, WeChat) from Google Play, and successfully detected 3 previously unknown bugs in Gmail (one parameter error) and Google+ (one UI update error and one lifecycle error). All of these bugs were detected in the latest versions at the time of our study, and have been reported to Google and got confirmed. However, these bugs have not been found by Monkey and Sapienz, while other testing tools, \eg, CrashScope~\cite{MoranVBVP17} and AppDoctor~\cite{HuYTY14} only consider 2 and 3 of these 8 enhancement cases, respectively.

\subsection{Enabling Exception Localization}
We find developers usually take days to fix a framework exception. Thus, automatically locating faulty code and proposing possible fixes are highly desirable. Our study can shed light on this goal.

\noindent{\emph{\textbf{Demonstration of Usefulness}}}. We have built a framework exception localization tool, ExLocator, based on Soot~\cite{soot}, which takes as input an APK file and an exception trace, and outputs a report that explains the root cause of this exception. It currently supports 5 exception types from UI update, Lifecycle, Index, and Framework Constraint errors (As Figure~\ref{fig:fix_efforts} shows, these errors are more difficult to fix). In detail, it first extracts method call sequences and exception information from the exception trace, and classifies the exception into one of our summarized fault categories, and then utilizes data-/control-flow analysis to locate the root cause. The report gives the lines or methods that causes the exception, the description of the root cause and possible fixing solutions, and closely related Stack Overflow posts. 
We applied our tool on total 27 randomly selected cases from Github, and correctly locates 25 exceptions out of 27 (92\% precision) by comparing with the patches by developers. By incorporating additional context information from Android framework (\eg, which framework classes use \texttt{Handler}), our tool successfully identified the root causes of the remaining two cases. 
However, all previous fault localization work~\cite{Sinha09,Jiang10,Mirzaei15,Wu14} can only handle general exception types.

\section{Conclusion}
\label{sec:con}
This paper conducts a large-scale analysis of framework exceptions in Android apps. We constructed a comprehensive dataset that contains 16,245 unique exception traces. After investigating 8,243 framework exceptions, we identified their characteristics, evaluated their manifestation via popular bug detection tools, and reviewed their fixes. Our findings enables several follow-up research. We demonstrated the usefulness of our findings by two prototype tools.

\section{Acknowledgements}
We appreciate the anonymous reviewers for their valuable
feedback.
This work is partially supported by NSFC Grant 61502170, NTU Research Grant NGF-2017-03-033 and NRF Grant CRDCG2017-S04.
Lingling Fan is partly supported by ECNU Project of Funding Overseas Short-term Studies,
Ting Su partially supported by NSFC Grant 61572197 and 61632005, and Geguang Pu partially supported by 
MOST NKTSP Project 2015BAG19B02 and STCSM Project 16DZ1100600. Zhendong Su is partially supported by
United States NSF Grants 1528133 and 1618158, and by a Google Faculty Research Award.

\clearpage
\bibliographystyle{ACM-Reference-Format}
\balance
\bibliography{sigproc} 

\end{document}